\documentclass{csmadapter}

\jvol{XX}
\jnum{XX}
\paper{8}
\jmonth{October}
\jname{IEEE CONTROL SYSTEMS}
\pubyear{2024}
\usepackage{algorithm}
\usepackage{algpseudocode}
\usepackage{amsmath, amsfonts, amssymb, nicefrac}
\usepackage{hyperref}
\usepackage{dirtytalk}
\usepackage{color}
\definecolor{darkgreen}{rgb}{0,0.5,0}
\usepackage[colorinlistoftodos]{todonotes}
\usepackage{soul}
\setulcolor{red} 
\setstcolor{red} 
\sethlcolor{yellow}

\newcommand{\Def}{\overset{\textbf{def}}{=}}
\newcommand{\rhomax}{\rho_{\max}}

\newcommand{\RR}{\mathbb R}
\newcommand{\set}[1]{\left\{#1 \right\}}
\newcommand{\totvar}[2]{\underset{#2}{T.V.}\left(#1 \right)}
\newcommand{\norm}[1]{\left\|#1\right\|}
\newcommand{\lb}{\left\{}
\newcommand{\rb}{\right\}}

\newcommand{\NN}{\mathbb{N}}

\newcommand{\Bv}{\mathbf v}
\newcommand{\Bu}{\mathbf u}

\DeclareMathOperator{\loc}{loc}

\newcommand{\AC}{\mathcal A_{\mathfrak c}}
\newcommand{\Riemann}{\mathcal R}

\begin{document}

\title{Hierarchical Speed Planner for Automated Vehicles
\stitle{A Framework for Lagrangian Variable Speed Limit in Mixed Autonomy Traffic}}

\author{
HAN WANG\textsuperscript{*}, 
ZHE FU\textsuperscript{*},
JONATHAN LEE\textsuperscript{*}, 
HOSSEIN NICK ZINAT MATIN\textsuperscript{*}, 
ARWA ALANQARY\textsuperscript{*}, 
DANIEL URIELI\textsuperscript{\textbardbl}, 
SHARON HORNSTEIN\textsuperscript{\textbardbl}, 
ABDUL RAHMAN KREIDIEH\textsuperscript{*}, 
RAPHAEL CHEKROUN\textsuperscript{*, \textsection, \textsection\textsection}, 
WILLIAM BARBOUR\textsuperscript{\textdaggerdbl}, 
WILLIAM A.~RICHARDSON\textsuperscript{\textdaggerdbl}, 
DAN WORK\textsuperscript{\textdaggerdbl},
BENEDETTO PICCOLI\textsuperscript{\textdaggerdbl\textdaggerdbl}, 
BENJAMIN SEIBOLD\textsuperscript{**}, 
JONATHAN SPRINKLE\textsuperscript{\textdaggerdbl}, 
ALEXANDRE M.~BAYEN\textsuperscript{*}, AND
{MARIA LAURA} {DELLE MONACHE}\textsuperscript{*}
}

\affil{
    *--University of California, Berkeley\\
    \textsuperscript{\textbardbl}--General Motors\\
    \textsuperscript{\textsection}--Mines Paris, PSL University\\
    \textsuperscript{\textsection\textsection}--Valeo Driving Assistance Research \\
    \textsuperscript{\textdaggerdbl}--Vanderbilt University\\
    \textsuperscript{\textdaggerdbl\textdaggerdbl}--Rutgers University-Camden\\
    \textsuperscript{**}--Temple University\\
}

\maketitle

\dois{}{}

\vspace{1em}
\begin{summary}

\noindent \summaryinitial{T}his paper introduces a novel control framework for Lagrangian variable speed limits in hybrid traffic flow environments utilizing \textit{automated vehicles} (AVs).
The framework was validated using a fleet of 100 connected automated vehicles as part of the largest coordinated open-road test designed to smooth traffic flow.
The framework includes two main components: a high-level controller deployed on the server side, named Speed Planner, and low-level controllers called vehicle controllers deployed on the vehicle side. The Speed Planner designs and updates target speeds for the vehicle controllers based on real-time \textit{Traffic State Estimation} (TSE)~\cite{seo2017traffic}. The Speed Planner comprises two modules: a TSE enhancement module and a target speed design module. The TSE enhancement module is designed to minimize the effects of inherent latency in the received traffic information and to improve the spatial and temporal resolution of the input traffic data. The target speed design module generates target speed profiles with the goal of improving traffic flow. The vehicle controllers are designed to track the target speed meanwhile responding to the surrounding situation. The numerical simulation indicates the performance of the proposed method: the bottleneck throughput has increased by 5.01\%, and the speed standard deviation has been reduced by a significant 34.36\%. We further showcase an operational study with a description of how the controller was implemented on a field-test with 100 AVs and its comprehensive effects on the traffic flow.

\end{summary}

\newpage

\chapterinitial {T}he advent of \textit{automated vehicles} (AVs) has the potential to revolutionize the field of transportation in the next decades, in optimizing traffic flow, reducing congestion, and improving the overall efficiency of the transportation system. The focus of this paper is to introduce a novel control framework specifically designed for managing variable speed limits in a hybrid traffic flow environment, one that combines both automated and non-automated vehicles. This framework hinges on the synergistic interplay between two key elements: server-side algorithms functioning as a centralized planner, tasked with handling heavy computational load, and vehicle-side algorithms operating as the executing agents, adhering to the targets set by the central planner. Our novel approach aims to enhance \textit{traffic state estimation} (TSE), to optimize the design of target speed profiles for large-scale AV platoons, and to reduce energy  and boost the throughput of the overall traffic system.

This paper details the design of the Speed Planner, a high-level algorithm that designs target speed profiles for vehicles in moving traffic, with the goal to smooth traffic waves. The system was validated on a fleet of 100 connected automated vehicles in the \textit{MegaVanderTest} (MVT), which was the largest coordinated open-road test designed to smooth traffic flow.
%
Overall, the Speed Planner tested in the MVT showcases capabilities in optimizing traffic flow. The full potential can be better realized with a holistic approach that considers both technological advancements and human-centric factors, like driver acceptance rate and social acceptance. 


\section{Introduction}

The control of \textit{large-scale multi-agent systems} (MAS) is a challenging problem in many applications, including transportation systems. In the context of large-scale control of traffic, the scalability of the control algorithms is a critical factor in their practical applicability. The motivation of this research is to provide an analysis and an optimization framework for a large-scale traffic control problem. 
\textit{Variable speed limit} (VSL) is a control technique that aims at modifying the speed limit on roads to improve safety and reduce congestion by optimizing traffic flow. The actuation of this type of control can be of two classes: 1) \textit{Eulerian VSL} where the actuation is at a fixed location and the speed limit changes at different time intervals; 2) \textit{Lagrangian VSL} that directly controls the individual vehicles by taking into account the interactions between vehicles, such as acceleration, deceleration, and changes in direction, and uses this information to calculate an optimal speed for each vehicle that can improve traffic flow and reduce congestion.
Prior works have investigated the effects of the Eulerian VSL \cite{papageorgiou2008effects, ADFFP98, HDSH05, HXZ07, HHSSV08, HHSS09, HH10, CPPM10, CDW11, FC12, YLLZ13, CVH13,delle2017traffic, GGK14}. The authors in \cite{papageorgiou2008effects} showed how dynamic regulations of speed on highways can significantly enhance traffic flow and reduce congestion. In \cite{Rama1999} the effects of weather-controlled speed limits and signs for slippery road conditions on driver behavior on the Finnish E18 test site were investigated.  
The paper \cite{hegyi2005} discusses the optimal coordination of variable speed limits and ramp metering in a freeway traffic network, where the objective of the control is to minimize the total time that vehicles spend in the network.
The Eulerian VSL approach primarily relies on traditional control infrastructure like roadside variable message signs to deliver the command, which leads to several significant drawbacks in operational implementation \cite{han2017}. Due to fixed locations, these signs struggle to provide flexible control according to the ever-changing traffic conditions. The SPECIALIST \cite{hegyi2005} field test analysis indicated that control failures were likely due to dynamic traffic conditions. Additionally, driver compliance with signs instructions is also a crucial concern in operational situations. Furthermore, deploying message signs is capital intensive~\cite{azin202280}, as they often necessitate a series of consecutive signs, along with costly infrastructure like gantries, over large areas. 
The Lagrangian VSL  has several advantages over Eulerian VSL. It can react more quickly to changing traffic conditions and can adjust the speed limit of individual vehicles based on their local situation \cite{han2017}.  Additionally, AVs also work as a probe vehicle for the traffic flow. They can provide a more accurate and comprehensive view of traffic conditions than traditional VSL systems, which rely on very limited sensor data. By leveraging the data collected by the AVs, VSL systems can gain a more comprehensive and up-to-date picture of traffic patterns and adjust speed limits more effectively and more timely.
In this work, we will focus on Lagrangian VSL control using connected AVs. In the following, whenever we talk about VSL, we refer to Lagrangian VSL.

A MAS is a network of interconnected autonomous agents that collaborate and communicate to accomplish complex tasks or solve problems. Distributed control refers to the decentralized nature of decision-making and coordination among these agents, which allows for greater scalability and robustness. In this context, MAS enables distributed control by facilitating information exchange, adaptation, and coordination among agents, leading to more efficient and resilient solutions in diverse domains  \cite{shukla2017hierarchical,  wang2015decentralized, li2021module}. In distributed control, each agent possesses its own processing power and uses communication to share information, enabling decisions based on distributed data. Distributed control can be categorized into two subclasses: Decentralized and Hierarchical control \cite{li2017designing, gessing1985two, wang2021enhanced}. In decentralized control, decision-making is evenly distributed among nodes, allowing local decisions based on individual states and information without a single entity having full control. Conversely, hierarchical control maintains a clear hierarchy in decision-making authority, where lower-level nodes must often align with or adhere to decisions made by higher-level nodes.


When it comes to VSL, distributed control has the potential to overcome the scalability challenges of centralized algorithms in large-scale traffic networks. 
%
%
A distributed VSL system leverages the computing power of many AVs to make local decisions which may be partially coordinated using data communication, thus
avoiding the dependence on a single central agent. 
%
This approach can alleviate the computational bottleneck and communication congestion associated with centralized VSL systems, making it easier to scale to larger and more complex road networks. In a distributed VSL system, each AV can collect and analyze data from its local observation. Based on this data, each AV can adjust its speed and communicate with nearby AVs and infrastructure to coordinate their target speed tracking and avoid collisions. By working together, the AVs can optimize traffic flow and reduce congestion on the road network, resulting in a safer and more efficient driving experience for all vehicles. In fact, prior works had proven that using each single vehicle as a traffic controller brings advantages from a traffic standpoint. This has been analyzed in field tests \cite{stern2018dissipation, delle2019feedback} and in simulations \cite{Wu2018}. Subsequently, this has advanced the theoretical framework both in modeling via multi-scale PDE-ODE models \cite{jode,Cicic2018, Cicic2020} and in control framework with a portion of literature focusing on understanding the effects of automated vehicles with intelligent speed controllers immersed in traffic \cite{giammarino2020traffic, bayen2022control, goatin2022interacting, daini2022centralized, Liard20231190, cicic2020coordinating, Cicic2019}.


In this work, we propose a hierarchical VSL control with centralized consensus, which involves a hierarchical structure in which a central server coordinates the decisions of multiple lower-level, vehicle controllers for mixed autonomy traffic systems, i.e., systems where AVs are immersed in human driven traffic. 
This approach 
%
overcomes
the scalability challenges of VSL and improves traffic, throughput and emissions. Our numerical simulations demonstrate the 
efficacy of 
%
our hierarchical VSL method:
bottleneck throughput increased by 5.01\%, system Miles Per Gallon (MPG) substantially rose by 15.41\%, fuel consumption was reduced by 34.14\%, and speed standard deviation was reduced by a significant 34.36\%. 
The real-world performance analysis of MVT, based on various visualizations and metrics, underscores the  effect of our approach on traffic flow in diverse traffic scenarios. Furthermore, we highlight the critical role of driver behavior in the overall system performance, emphasizing the potential challenges of social acceptance when AVs operate at speeds optimized for traffic flow, but divergent from human-driven vehicle speeds.
Adaptive speed planning in response to real-time traffic conditions, and, the need for effective communication strategies to bridge the gap between AV operations and human driver expectations that we saw from MVT, offer valuable guidelines for the design and deployment of traffic management systems in an era of increasing AV penetration.  

\section{Methodology}
\label{s:methodology}

\begin{pullquote}
    The proposed speed planner introduces a hierarchical framework for variable speed limit, specifically designed to feed in-vehicle controllers in mixed traffic flow.
\end{pullquote}
This section introduces a hierarchical framework for the VSL control system, specifically designed for mixed-autonomy traffic flow. 

\begin{figure}[h!]
    \centering
    \includegraphics[width=12cm]{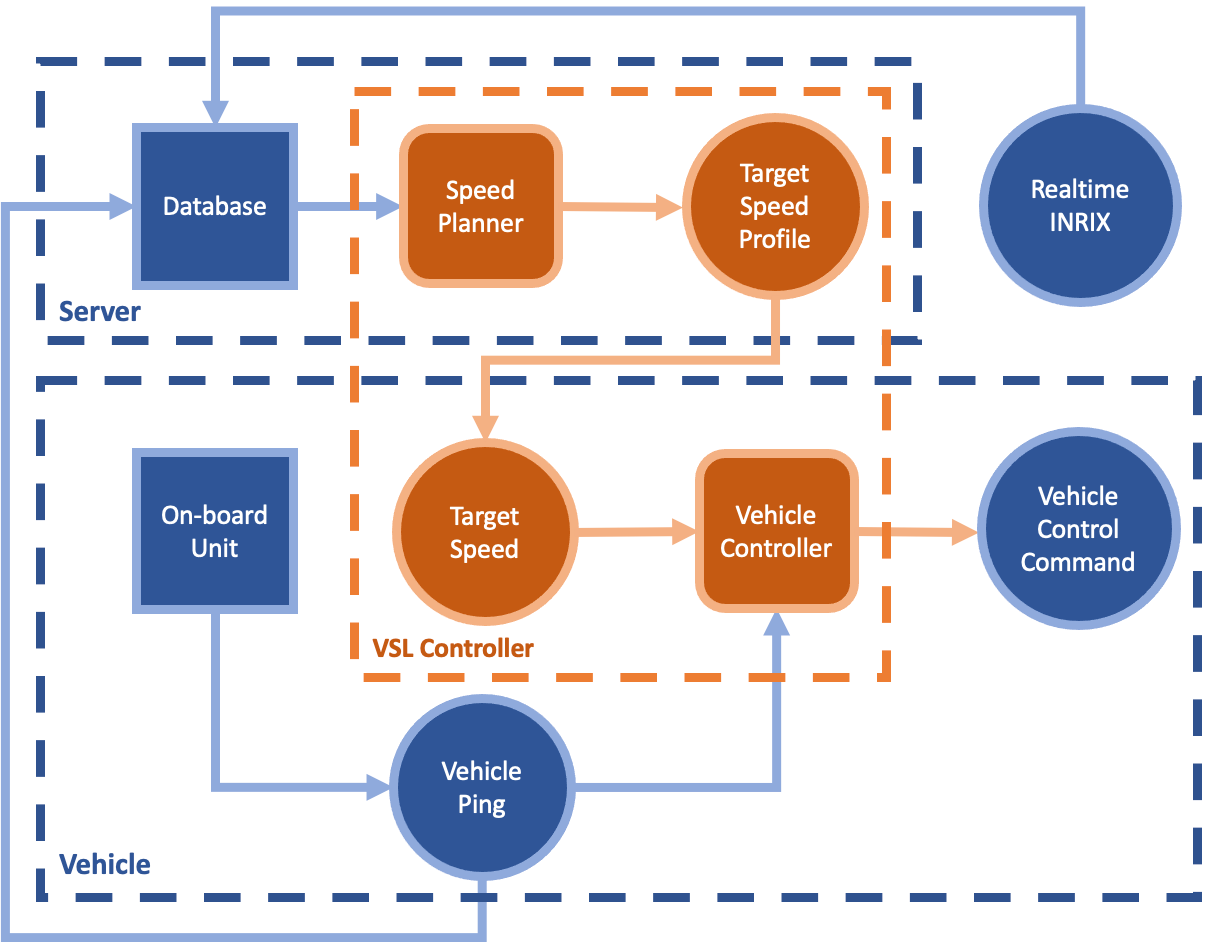}

    \caption{Hierarchical framework of the proposed VSL system: The Speed Planner module fetch inputs from the database to generate real-time target speed profile. Vehicle controllers get assigned targets speed via an API, together with local observation collected by on-board unit, as the input to decide the instant vehicle control command.} 
    
    \label{fig:framework}
\end{figure}

\subsection{Hierarchical Framework}

The hierarchical framework, as illustrated in \textbf{Fig \ref{fig:framework}}, forms the foundation of our advanced traffic management system. Designed to amalgamate both macroscopic and microscopic TSE seamlessly, it generates an optimal target speed profile for the controlled vehicles. The primary objective of this framework is to harmonize traffic flow, alleviate congestion, and bolster road safety by harnessing real-time data and sophisticated algorithms. To do so, we build a control architecture (see \textbf{Fig \ref{fig:framework}}) that leverages traffic data and prediction to find the optimal speed profile based on  different traffic conditions. The rest of the section is devoted to the description of each element of the control architecture. We first give a general overview of each element and then we will describe in details each one of them. 

\paragraph{\textbf{Data Collection and Integration}}
Central to our system is a comprehensive database situated on the server side. This database is instrumental in accumulating and processing traffic data from two main sources:

1. \textbf{Macroscopic traffic state from INRIX}. INRIX \cite{cookson2017inrix} feeds our system with macroscopic data, presenting average speed metrics for predefined road segments, updated every minute. A deeper dive into the INRIX data is available in the sidebar "External Data Source".

2. \textbf{Microscopic observations from controlled vehicles}. Offering a more detailed view, controlled vehicles furnish our system with a microscopic snapshot of the traffic milieu. These vehicles, equipped with onboard detectors, capture data at 0.1-second intervals and relay this information to the central server every second. Each data transmission includes the status of the transmitting vehicle and its perception of its immediate surroundings. Data from GPS as well as vehicle state are gathered directly from vehicle sensors through Libpanda \cite{bunting2021libpanda}, which uses the strym package \cite{bhadani2022strym} to decode vehicle data from proprietary formats at runtime. Using custom software bridges \cite{elmadani2021can}, a web service that runs as part of vehicle middleware \cite{nice2023middleware} publishes data to the central server on which the Speed Planner is hosted, at approximately 1 Hz.

\paragraph{\textbf{TSE Enhancement}}
To refine the accuracy and reliability of the traffic data we created a TSE Enhancement algorithm that estimates the traffic state based on INRIX and vehicle data. The TSE Enhancement is based on: 

1. \textbf{Prediction Module}. Historical INRIX data is pulled and used  
 as the primary input for this module. This predictive approach ensures that our system remains proactive, anticipating traffic patterns and counteracting INRIX latency.
   
2. \textbf{Observation Fusion}. Vehicle observations from the preceding minute are integrated into the macro TSE. By fusing this granular data with the broader INRIX predictions, we derive a lane-specific TSE, ensuring a comprehensive understanding of both macro and micro traffic dynamics.

\paragraph{\textbf{Target Design}}
The data extracted from the TSE Enhancement is used to generate the target speed profile which is the heart of our traffic management strategy. Different types of congestion (bottleneck or shockwave) will be taken into account and the target speed profile will reflect the best strategy for each case. The following modules are the main elements of the target speed design:

1. \textbf{Kernel Smoothing} \cite{fu2023cooperative}. To derive a consistent and actionable TSE, the fused data undergoes a smoothing process using a forward average kernel. This step provides robust countermeasures for shockwaves, which are high-density traffic waves that propagate towards upstream, often resulting from sudden traffic disruptions and extra fuel consumption. By addressing these shockwaves, we ensure that transient anomalies in the data do not skew the overall traffic understanding.

2. \textbf{Buffer Design}. Upon detecting a standing bottleneck via the bottleneck identification module, a buffer segment is crafted within the smoothed TSE. This design ensures that vehicles are provided with speed recommendations that account for standing bottlenecks, promoting the efficiency of traffic flow.

3. \textbf{Optimal Planner}. In scenarios devoid of bottlenecks, the smoothed TSE itself is utilized as the target speed profile. This profile is then broadcast, guiding individual vehicle controllers to make decisions at a frequency of  0.1 seconds. The control decisions are represented as a two-element array $[V_{ACC}, X_{gap}]$, where $V_{ACC}$ denotes the speed setting, and $X_{gap}$, ranging between $\{0,1,2\}$, indicates the gap setting for the \textit{adaptive cruise control} (ACC) system \cite{pananurak2009adaptive}.

In the following sections, we will delve deeper into each facet of this hierarchical framework, shedding light on the methodologies, algorithms, and strategies employed to ensure optimal traffic management.

\begin{figure}[h!]
    \centering
    \includegraphics[width=12cm]{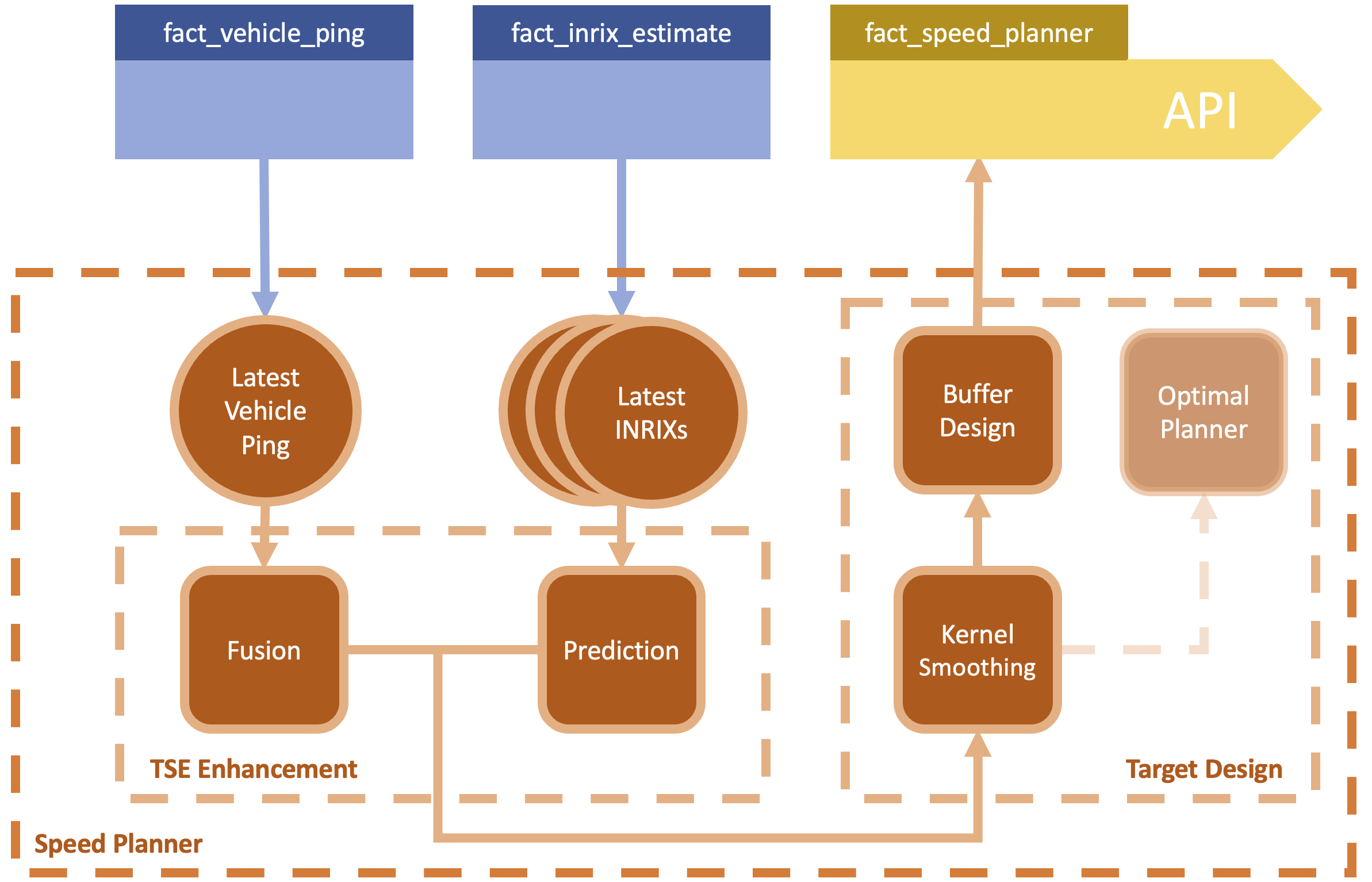}

    \caption{Data pipeline for the Speed Planner: 1.For each update, past INRIX data from the database are fetched as the input of the prediction module. 2. Fetch the vehicle observations of the previous 1 minute. Fuse the INRIX prediction with vehicle observation to obtain the lane level TSE. 3. Smooth the obtained lane level TSE with the forward average kernel. 4. Input the smoothed TSE into the bottleneck identification module. 5. If there is any standing bottleneck identified, design the corresponding buffer segment in the smoothed TSE as the target speed profile. Else use the smoothed TSE as the target speed profile. 6. Publish.}
    
    \label{fig:planner}
\end{figure}


\subsection{TSE Enhancement}\label{s:tse_enhancement}

\begin{sidebarauthor}{External Data Source: Average Segment Speed from INRIX}{Han Wang}

\setcounter{sequation}{0}
\renewcommand{\thesequation}{S\arabic{sequation}}
\setcounter{stable}{0}
\renewcommand{\thestable}{S\arabic{stable}}
\setcounter{sfigure}{0}
\renewcommand{\thesfigure}{S\arabic{sfigure}}

\sdbarfig{\includegraphics[width=19.0pc]{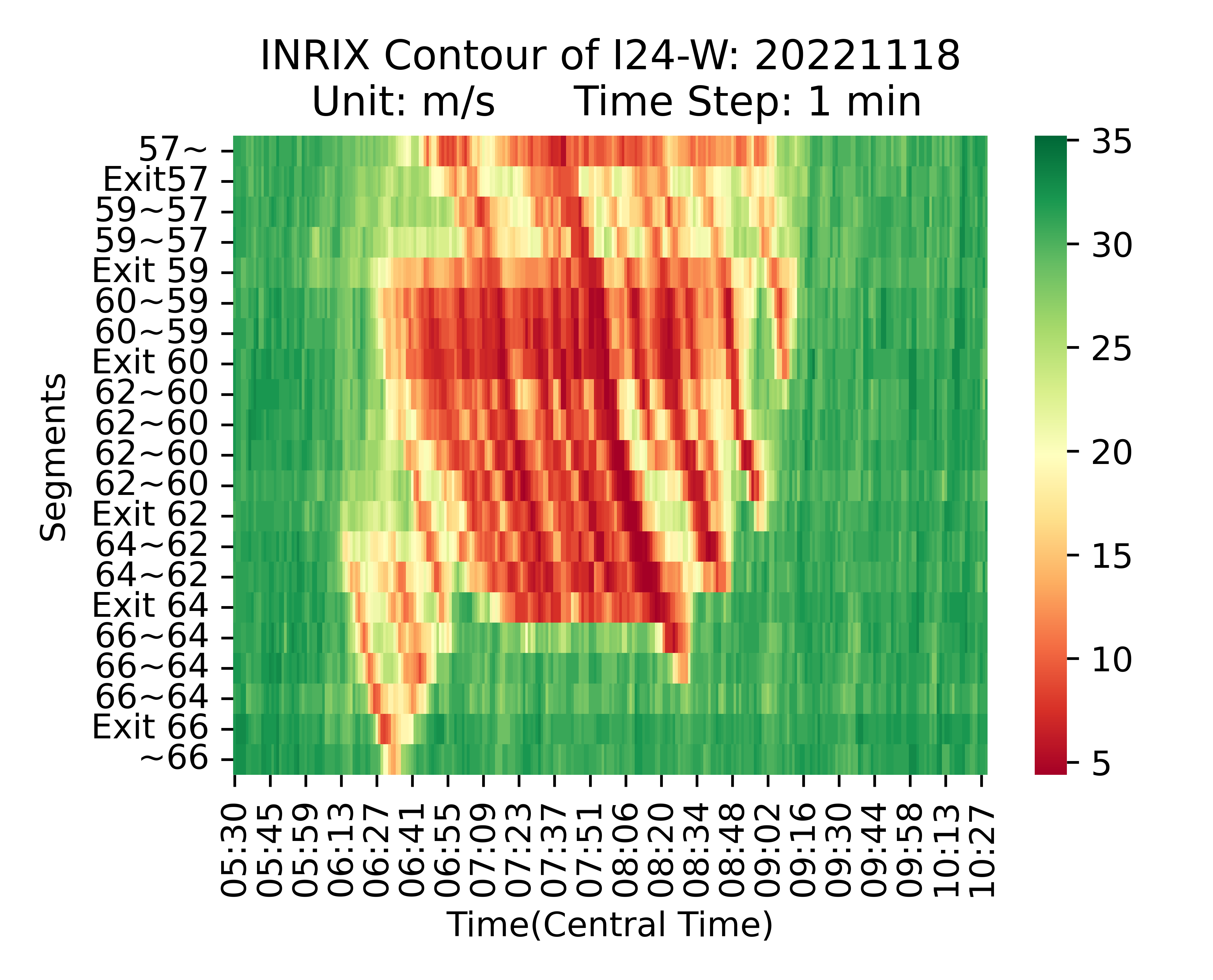}}{Speed Heatmap \label{fig:inrix1} of I-24 Westbound on November 18, 2022: the color of each cell indicates the average speed of vehicles in the corresponding time-space window. The traffic flow moves from the bottom towards the top of the figure. A standing bottleneck can be observed at Exit 59.}

The API provided by INRIX delivers a real-time assessment of aggregated velocity for distinct sections of the highway infrastructure. This data undergoes regular updating as new information surfaces, thereby facilitating an instantaneous representation of prevailing traffic conditions. As illustrated in Figure \ref{fig:inrix1}, a heatmap shows the average velocity of vehicles along I-24 where the experiment was conducted, updated at an interval of one minute. 

Geographical segmentation underpins the operational modules of INRIX data. This implies that highways are separated into various segments, and the average speed is calculated separately for each segment. The segment size can vary, chosen to yield an appropriate level of detail. As an example, a segment could be the section of a highway between two interchanges. In correspondence of each highway segment, INRIX gathers data from every vehicle located in the segment and using these, INRIX computes the average speed. In this paper, we adopt the following notation to represent the discrete data we fetched from the INRIX API:
\begin{equation}\label{E:inrix_seg}
\{(x_j,\Bar{v_j}): j \in \mathcal J\},
\end{equation}
where $\mathcal J$ represents the collection of INRIX road segments, and for any $j \in \mathcal J$, $x_j$ is the postmile representing the center of the specific road segment $j$ and $\Bar{v_j}$ is the average speed of the corresponding road segment.

While INRIX's data is invaluable for many applications, especially in traffic analysis, there's an inherent latency in the data when considering real-time control tasks. Given the vast coverage and the depth of analysis they provide, a latency of 3 minutes is commendably low. However, for our specific control task, even such a minor delay can be significant. This sensitivity to latency underscores the need for predictive models, like the TSE enhancement module implemented in this study, to anticipate and adjust for these delays, ensuring more accurate real-time TSE.

\end{sidebarauthor}

The TSE enhancement module consists of a prediction and a fusion module. The purpose of the prediction module is to mitigate the effects of the latency of INRIX real-time data. The fusion module further improves the TSE by introducing real-time observations of the controlled vehicles of our system to obtain lane-level TSE with higher time-space resolution. 

\begin{figure*}[p]
    \centering
    \includegraphics[width=12cm]{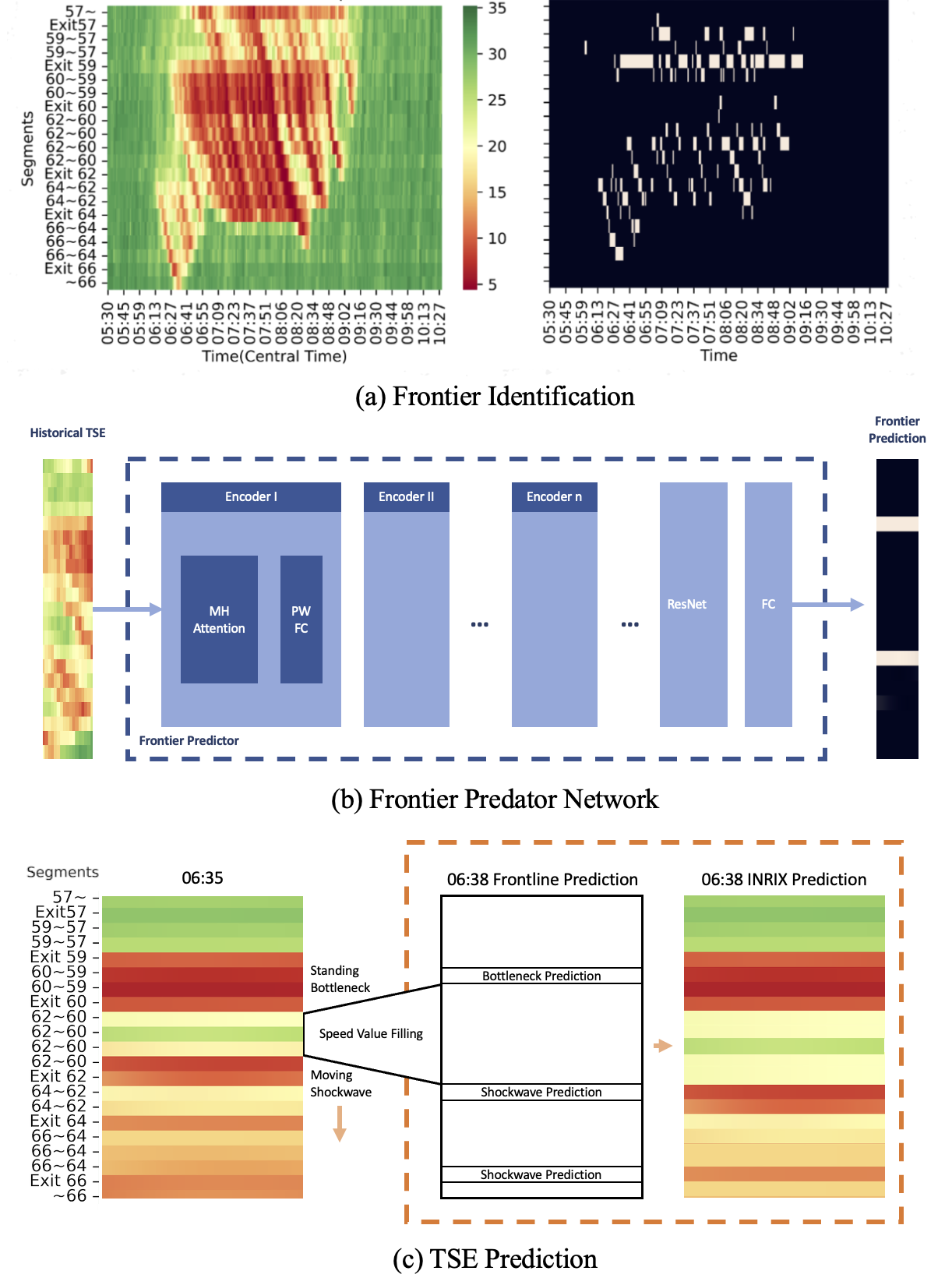}

    \caption{Prediction Module for TSE Enhancement: (a) INRIX heatmap on November 18th, 2022 (Left) and congestion frontier identification result (Right); (b) Network architecture of frontier predictor; (c) Speed filling to obtain TSE prediction.}
    
    \label{fig:prediction}
\end{figure*}

The TSE enhancement module aims to refine the accuracy and reliability of our traffic state predictions. This enhancement is achieved through a two-phase procedure, as depicted in Figure \ref{fig:prediction}.

\paragraph{\textbf{Prediction Module}}

The prediction module aims to alleviate the effects of the 3-minute latency in INRIX data by forecasting the traffic condition 3 minutes ahead. The core of our prediction module is a neural network that incorporates a self-attention layer \cite{vaswani2017attention}. This network is trained to identify and predict the movement of congestion frontiers within the INRIX data. 

\textbf{Training Data}. The network is trained on morning INRIX data spanning from January 01, 2020, to October 30, 2022. This dataset is labeled based on the PeMS bottleneck identification procedure \cite{chen2004systematic}. In a nutshell, a position will be identified as a congestion frontier if its average speed is monotonically lower than its adjacent downstream and upstream positions, and the rate of speed reduction between these positions exceeds a predefined threshold. Figure \ref{fig:prediction} (a) show how the frontier identification on November 18, 2022 INRIX data of  works.

\textbf{Data Input}. For each update, the INRIX data from the preceding 6 minutes, tagged with the time of day, is input into the self-attention layer.

\textbf{Network Output}. The network produces a binary vector, mirroring the length of the INRIX input. Each binary element indicates whether the corresponding INRIX segment is a congestion frontier, as illustrated in Figure \ref{fig:prediction} (a).

\textbf{Network Structure}. The Transformer model we employ consists of an encoder stack (see Figure \ref{fig:prediction} (b)), with each encoder comprising a multi-head self-attention mechanism and a position-wise fully connected feed-forward network. The self-attention mechanism allows the model to focus on different parts of the input TSE, enabling it to recognize congestion patterns over varying time scales and different spatial contexts. Layer normalization and residual connections are also incorporated to facilitate training and improve convergence.

\textbf{Input Embedding}. The INRIX data, along with the time of day, is embedded into a high-dimensional space using learned embeddings. Additionally, positional encodings are added to these embeddings to provide the model with information about the order of the data segments.

\textbf{Training Procedure}.
We employed the following items for the training of frontier predictor.
(i) \textit{Loss Function}. We employ a binary cross-entropy loss, given our task is to predict whether a segment is a congestion frontier (binary classification).
(ii) \textit{Optimizer}. The Adam optimizer \cite{kingma2014adam} is used with a learning rate that follows a warm-up and then decay schedule.
(iii) \textit{Regularization}. Dropout is applied between layers to prevent overfitting, especially given the potential for noise in traffic data.
(iv) \textit{Batching}. Mini-batch training is employed, with batches constructed by grouping segments of similar lengths to minimize padding. 
(v) \textit{Model Evaluation and Validation}. The model is evaluated using a held-out validation set, ensuring that it generalizes well to unseen data. Key metrics include precision, recall, and the F1 score, focusing on the model's ability to accurately identify congestion frontiers.

\textbf{Congestion Frontier Types}. By leveraging the historical data access of our centralized Speed Planner, we introduce criteria to discern the nature of the congestion frontier. Continuous recognition of a location as a congestion frontier marks it as a standing bottleneck. Conversely, if a frontier consistently moves upstream at a relatively uniform speed, it's identified as a shockwave. Recognizing these congestion types aids in tailoring countermeasures in the target speed profile.

\textbf{Frontier Tracking}. The recognized congestion frontiers serve as anchor points for mapping speed values from the latest INRIX data. As showcased in Figure \ref{fig:prediction} (c), each wave frontier is tracked across successive frames. The speed arrays are then adaptively mapped to these tracked slots, undergoing interpolation and compression to accommodate varying interval lengths.

\paragraph{\textbf{Fusion Module}}

To achieve a higher spatial resolution for lane-level TSE, we further segment the INRIX data into smaller sub-segments. Once the INRIX prediction is obtained, it is merged with real-time observations from controlled vehicles. For each target speed profile update interval, we obtain 30 ping records from each vehicle. These records compute each vehicle's average speed over the past update interval, subsequently overwriting the TSE of the corresponding sub-segment. Given that vehicle lane data is included in the pings, a distinct TSE is crafted for each lane by amalgamating corresponding vehicle observations with the overarching INRIX prediction. We opt to directly overwrite INRIX data with our controlled vehicle data due to the transparent and controllable nature of our system.

\paragraph{}

The output of the TSE Enhancement will be a series of discrete data pairs $(x_j, \Bar{v}_j)$, $j \in \mathcal J$ that link a location in the road with speed values; see Equation \eqref{E:inrix_seg}.

%



\begin{pullquote}
Hierarchical coordinated speed planning for CAVs can perform even in the absence of V2V coordination using commercial traffic data.
\end{pullquote}
\subsection{Target Design}
\label{sec:speed_planner}
The enhanced TSE is used in the design module to generate the target speed profile for controlled vehicles. This section introduces the major function modules, including kernel-based smoothing, learning-based buffer design, and optimization-based planner. The kernel smoothing processes the enhanced TSE at each time step using a chosen kernel to improve the fuel consumption caused by the shockwave in a high-density traffic flow. 

The buffer design utilizes \textit{reinforcement learning} (RL) to form a buffer area upstream of the standing bottleneck with the goal of improving throughput at the bottleneck. The target speed suggested by the RL will be employed in the mathematical model of traffic, represented by a strongly coupled \textit{partial and ordinary differential equation} (PDE-ODE). The outcome of this mathematical model is an identification of traffic density $(t, x) \mapsto \rho(t, x)$. The kernel smoothing then will receive this information for learning the velocity of the next time step. Furthermore, an optimal planner will be introduced which provides an alternative optimization-based benchmark to the learning-based approach. We now look into each of these approaches in more detail.


 \subsubsection{Kernel Smoothing} \label{S:kernel_smoothing}
 The purpose of kernel smoothing is to create a speed profile that  minimizes the adverse effects of shockwaves and fuel inefficiency.
In other words,  kernel smoothing generates a speed profile that optimizes fuel efficiency by predictively and adaptively reacting to the downstream traffic state at each fixed time step.

The proposed approach adopts and extends prior heuristic approaches on traffic flow harmonization~\cite{asadi2010role,cui2017stabilizing,stern2018dissipation,kreidieh2022learning}, which posit that traffic may be homogenized near its desirable \emph{uniform} driving speed by operating a subset of vehicles near accurate predictions of said speed.

The desirable uniform driving speed must be achieved without shared communication between adjacent vehicles, since in mixed-autonomy settings not every vehicle may be able to communicate. Instead, we rely on enhanced TSE data to synchronize the driving speeds of automated vehicles. In particular, vehicles are assigned target speed profiles contingent on traffic state information, which is shared and common among all AVs. 

At any fixed time step $t$, the desired speed profile $\Bv:\RR_+ \times \RR \to \RR_+$ is extracted from the enhanced TSE utilizing kernel methods. First, we preprocess the sparse TSE data by interpolating the discrete data pairs $(x_j, \Bar{v_j})$ for $j \in \mathcal J$ to a continuous speed profile $(t, x)  \in \RR_+ \times \RR \mapsto v(t, x)$, as an approximation of the average speed of a higher granularity traffic at a position $x$ and at time $t$. Then, for any fixed time $t = t_\circ$ we obtain the desired speed by applying a kernel function $K(\cdot)$ at a position $x = x_\alpha$: 
\begin{equation}
    \Bv(t_\circ, x_\alpha) = \frac{\int_{x=x_\alpha}^{x_\alpha+w} K(x_\alpha, x) v(t_\circ,x)dx}{\int_{x=x_\alpha}^{x_\alpha+w} K(x_\alpha, x)dx}, \label{eq:desire_kernel}
\end{equation}
where $w$ is the width of the estimation window. Many different kernel functions, such as Gaussian kernel, Triangular kernel, Quartic kernel, Uniform kernel etc., can be chosen. For the purposes of this paper, we consider a uniform kernel, the simplest of such mapping. The desired speed profile at a position $x_\alpha$ is accordingly defined as:
\begin{equation}
    \Bv(t_\circ, x_\alpha) = \frac{\int_{x=x_\alpha}^{x_\alpha+w} v(t_\circ, x)dx}{w}. \label{eq:desire}
\end{equation}
 
For human drivers, when they observe a gap between their vehicle and the one preceding, they tend to accelerate to close the distance. Our proposed desired speed profile aims to slow down \emph{in advance}, although not excessively, to create a gap from the preceding vehicle. This approach takes into account the information provided by the TSE, which indicates the presence of congestion in the nearby downstream area. The proposed desired speed profile is adaptive to traffic states and offers relative robustness, as it only requires one parameter, $w$, to tune.



\subsubsection{Buffer Design}  
The broad idea of designing a buffer area is to minimize congestion in the downstream traffic due to the presence of a standing bottleneck. 


From the perspective of speed planning, the problem caused by congestion can be interpreted as a reduction in system efficiency due to the uneven distribution of traffic density in the time-space domain: the bottleneck causes a throughput reduction due to the high density of queues area, and shockwaves cause the waste of energy due to the propagation of high-density waves. The Speed Planner achieves traffic flow efficiency by navigating the controlled vehicles to regulate the traffic flow and achieve a uniform density distribution. 

\begin{figure}[h!]
    \centering
    \includegraphics[width=12cm]{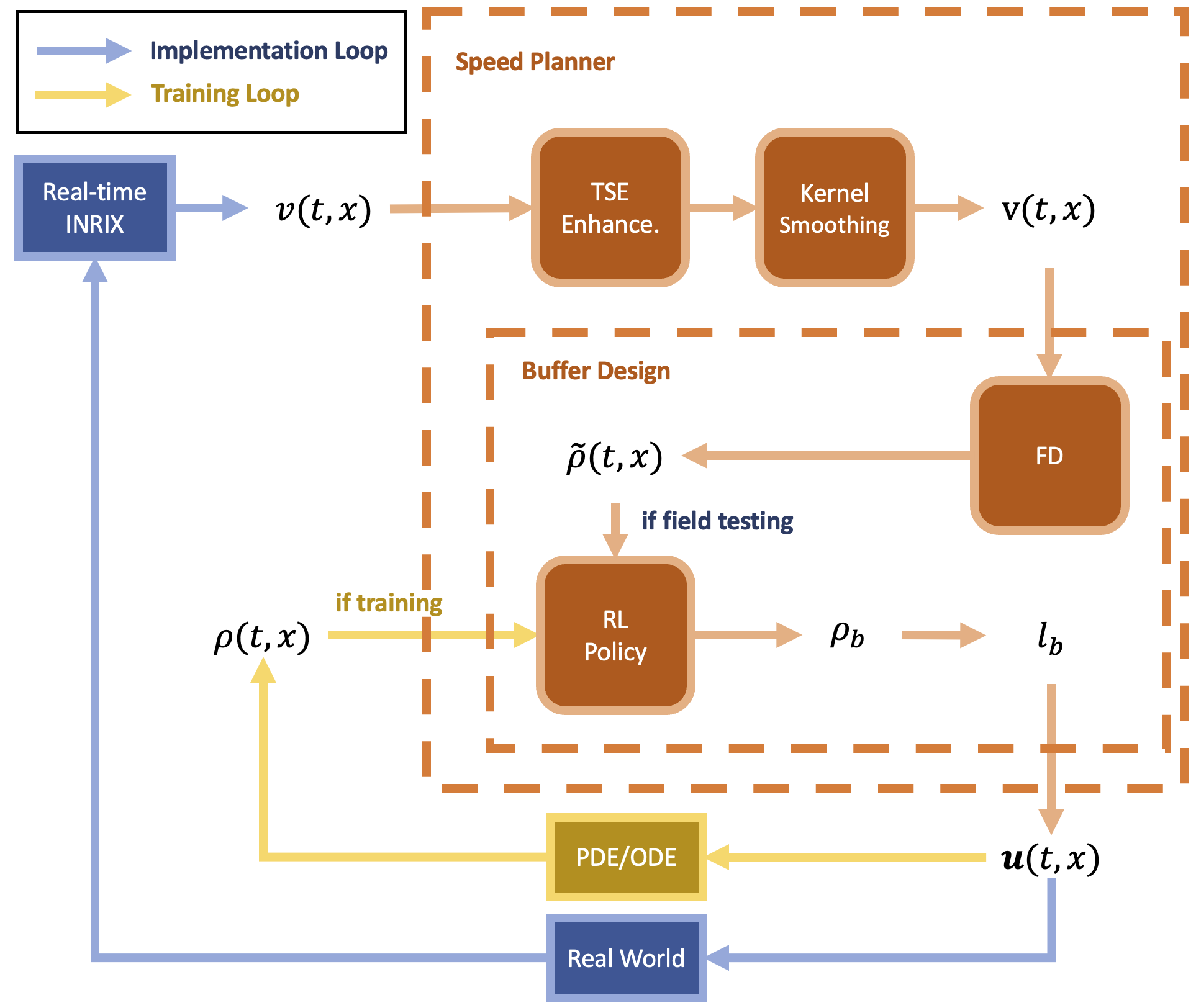}
    \caption{Training and field testing pipeline of the buffer design module.}
    \label{fig:buffer_design_training}
\end{figure}

\begin{figure}[h!]
    \centering
    \includegraphics[width=12cm]{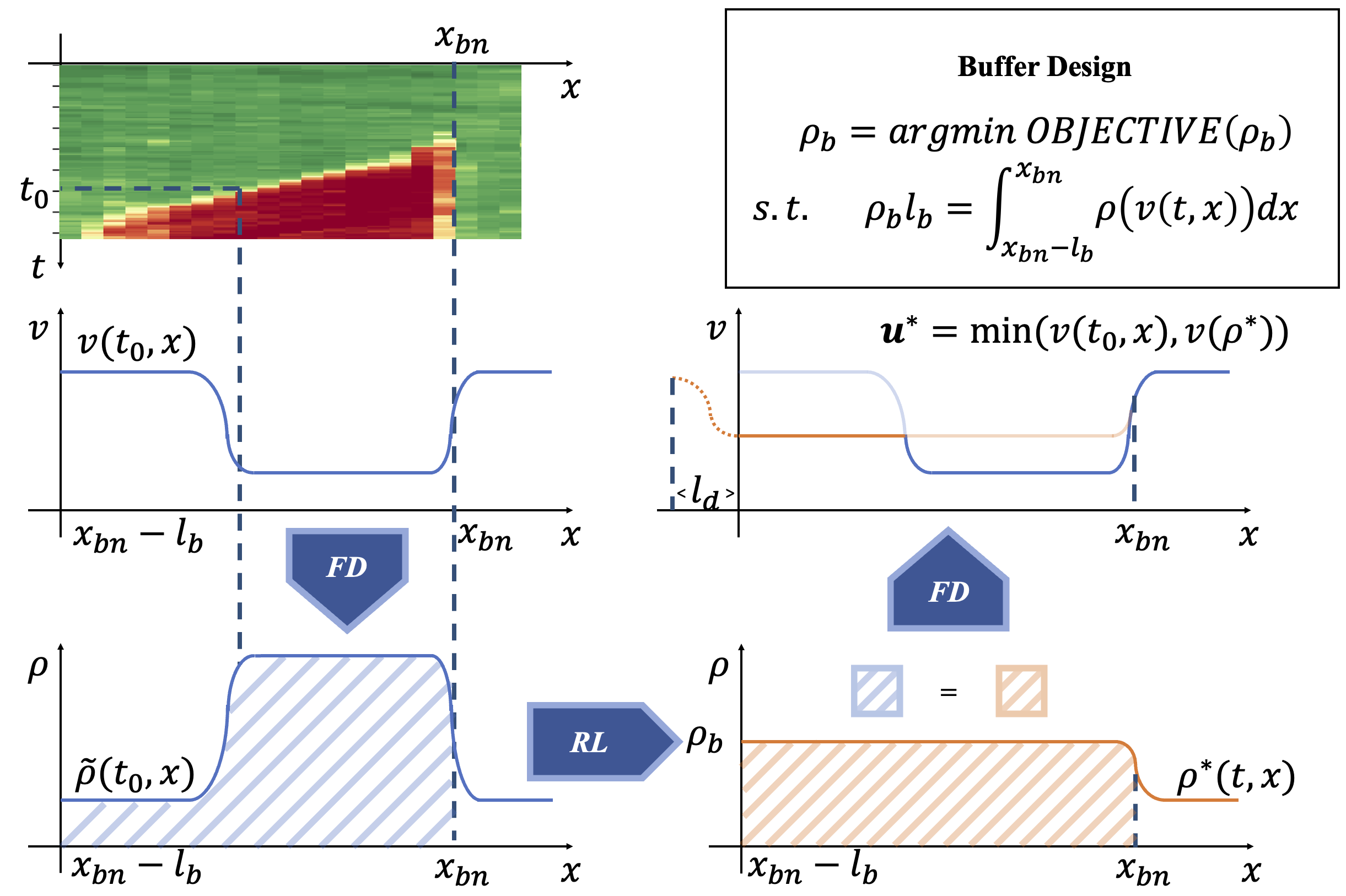}
    \caption{Obtain target speed profile: once a standing bottleneck is identified, the speed profile $v(t,x)$ will be converted to the density profile $\tilde\rho(t,x)$ by using a calibrated FD. RL policy selects the desirable density for the buffer area $\rho_b$ based on $\tilde\rho(t,x)$, which is the critical parameter to determine a desirable density profile $\rho^*(t,x)$. The target speed profile is obtained by converting $\rho^*(t,x)$ to the speed profile and taking the minimum value at each position.}
    \label{fig:buffer_area_construct}
\end{figure}

In this section, we detail the buffer area design process from a theoretical standpoint. 
We consider the interval $\mathcal I\subset \RR$ as the region of interest. In addition, we consider a subregion $\mathcal I_c \subset \mathcal I$ as a congested area (see Figure \ref{fig:buffer_area_construct}). The idea is to determine the controlled vehicle target speed at a time step $t_\circ$, denoted by $\Bu(t_\circ, x)$, such that the density $\rho(t, x)$ for $x \in \mathcal I_c$ and $t \ge t_\circ$ is distributed uniformly through the region $\mathcal I$. Determining the controlled vehicle target speed will be done in the following steps: 
(i) Designing a target speed $\Bu(t_\circ, x)$, given the input $\Bv(t_\circ, x)$ from kernel smoothing step, 
(ii) identifying the density $(t, x)\in \RR_+ \times \mathcal I \mapsto \rho(t,x)$, given the target speed of the controlled vehicle, employing a strongly coupled PDE-ODE model of traffic flow, 
(iii) Evaluation step in which using the density, the speed profile will be updated by smoothing kernel. 
Figure  \ref{fig:buffer_design_training} indicates the data pipeline of the Buffer design module in the training loop and the MVT implementation loop separately, which will be detailed in the following of this section. A noteworthy difference is the TSE representation: in the training loop, the environment mechanism modeled by PDE/ODE takes the target speed profile as input, and identifies the density profile $\rho(t,x)$ as the next step input. In the field test implementation, given that the available TSE only provides the estimation of the average speed of road segments, a mapping from speed to density, i.e. \textit{fundamental diagram} (FD) in traffic flow theory~\cite{daganzo1997fundamentals,greenberg1959analysis,newell1961nonlinear}, is modeled to estimate the traffic density at different locations of the road. The FD model converts the estimation of the speed profile into the density profile of each road segment, i.e.  $v(t,x) \overset{FD}{\mapsto} \tilde \rho(t, x)$; see Figure \ref{fig:buffer_design_training}. The section "\nameref{sec:field_test_analysis}" provides the evaluation of the approximation performance of the FD calibrated in MVT.


\paragraph{\textbf{Designing the Speed Profile}} 
Following the TSE enhancement and kernel smoothing, we find the congestion frontiers. If a standing bottleneck is identified, the lane level TSE will be considered as an input for the buffer design module to generate a target speed profile. The spirit of buffer design is to inform the vehicles to start braking in advance when driving into congestion. 

The buffer area is determined by two critical parameters: desired density $\rho_b$ and buffer length $l_b$. From the density profile, the continuous upstream sections of the bottleneck whose density exceeds a preset threshold are identified as a queue area, as shown in Figure \ref{fig:buffer_area_construct}. To distribute the density evenly in the queue area upstream, $\rho_b$ and $l_b$ should be designed to satisfy the following equality: 
\begin{equation}
\label{eq:l_b}
l_b(t_\circ) = \tfrac{1}{\rho_b(t_\circ)} \int_{x_{bn}-l_b}^{x_{bn}}  \rho(t_\circ, x)dx
\end{equation}
where $x_{bn}$ denotes the location of the standing bottleneck, and the desired density $\rho_b$ is the control variable for the designed optimizer. If $\rho_b$ is controlled then, the length of the buffer area $l_b$ could be obtained by recursively solving equation \eqref{eq:l_b}. In the implementation phase, the integrand will be replaced by $\tilde \rho(t_\circ, x)$ (see Figure \ref{fig:buffer_design_training}).

A deceleration area connects the buffer area with the non-congested areas upstream to ensure that traffic decelerates smoothly while driving into the buffer area and prevents secondary shockwaves from being generated in this process. The deceleration area is designed in the time-speed domain and determined by two elements: the desired constant decelerate rate $a_d$ and the deceleration area length $l_d$ that can be calculated by solving 
\begin{equation*}
l_{d}(t_\circ)= \tfrac{1}{2a_{d}} \lb(\Bv(t_\circ, x_{bn}-l_{b}-l_d))^2- (\Bv(t_\circ, x_{bn}-l_b))^2 \rb .
\end{equation*}
In MVT, we designed the controller utilizing the RL algorithm with a PDE-ODE based environment mechanism to dynamically update the parameters of target speed profiles according to real-time traffic conditions.

To apply the RL algorithm, we model the problem into a \textit{Markov decision process (MDP)} problem \cite{puterman2014markov}. A finite MDP problem can be represented by a tuple $\mathcal{P}(\mathcal S,\mathcal A,P,R)$. For agents in MPD at any time step $t_\circ$, applying an action $a(t_\circ) \in \mathcal A$ to the observed state $s(t_\circ)\in \mathcal S$ results in a new state $s(t_\circ+1)$ with probability $P(s(t_\circ+1)|(s(t_\circ),a(t_\circ)))$ and a reward function $r(t_\circ)=R(s(t_\circ),a(t_\circ),s(t_\circ+1))$. After updating the current state to $s(t_\circ+1)$, the process will be repeated until the end criteria of the episode are met. In our problem setting, the state space $\mathcal S$ is the enhanced and smoothed TSE defined as the speed profile $\Bv(t_\circ,x)$; the action space $\mathcal A$ is the domain of the definition of the desired density $\rho_b$, which corresponds to the target speed $\Bu(t_\circ, x)$ as shown in Figure \ref{fig:buffer_area_construct}. The policy then selects one value from the action space $\mathcal A$ which will be used to design the buffer area in the target speed profile; the reward function is the throughput at the bottleneck. The transfer of state $\mathcal{P}(\mathcal S,\mathcal A,P,R)$ is obtained by solving the underlying mathematical model of traffic (see Equation \eqref{E:main} below). The policy $\pi: s\in \mathcal S \to \mathcal A$ is a mapping from the state space $\mathcal S$ to the action space $\mathcal A$. The objective of the MDP problem is to find the optimized policy $\pi^*$ to maximize the total reward (expectation) and calculate the target speed $\Bu(t_\circ, x)$ based on the observed state in a fixed time step $t_\circ$. 




\paragraph{\textbf{The Mathematical Model of Traffic Flow and Density Prediction}}
The behavior of the traffic flow in the presence of controlled vehicles, will be modeled using a mathematical (dynamical) model comprising PDE-ODE. Such problems are studied both in engineering communities \cite{lebacque1998introducing, leclercq2004moving, giorgi2002traffic} as well as from a mathematical point of view \cite{lattanzio2011moving, delle2014scalar, delle2017stability, liard2019well, liard2021entropic}. 
In particular, we model traffic dynamic as follows:

\begin{equation}\label{E:main}
\left\lbrace
\begin{array}{ll}
    \rho_t + [f(\gamma, \rho)]_x = 0 &  \scriptstyle{(t, x) \in \RR_+ \times \RR}\\
      \rho(0, x) = \rho_\circ(x)   & \scriptstyle{ x \in \RR} \\
      \dot y_i(t) = w(y_i(t),\rho(t, y_i(t)+)) & \scriptstyle {i \in [1,\dots, N]}\\
      y_i(0) = y_{\circ,i} & \scriptstyle{i \in [1,\dots, N] }\\
      f(\rho(t, y_i(t))) - \dot y_i(t) \rho(t, y_i(t)) \le F_\alpha(y_i, \dot y_i)  & \scriptstyle{ i \in [1,\dots, N] }
\end{array}
\right. 
\end{equation}

The PDE in \eqref{E:main} is simply the classical LWR model \cite{LW55, R56} augmented with a discontinuous flux.  We divide the highway $\mathcal I$ into $M + 1$ subregions $I_m$, for $m \in \set{0, \cdots, M}$. The function $\gamma(x)$ represents the maximal speed $V^{(m)}_{\max}$ in each region $I_m$, $m \in \set{0, \cdots, M}$ of the road, and it is considered to be piecewise constant; i.e., $\gamma(x) = V^{(m)}_{\max}= \gamma_m$ for each $m$. 
\begin{remark}
   Note that during MVT the speed limit remains constant, i.e., $\gamma(x) = V_{\max}$ for all $x \in \mathcal I$. However, in this description  we adopt a more general analytical approach. 
\end{remark}

Function $(t, x) \in \RR_+\times \RR \mapsto \rho(t, x)\in [0, \rho_{\max}]$ denotes the density function. The flux function $f: \set{\gamma_\circ, \cdots, \gamma_M}  \times [0, \rho_{\max}] \to \RR_+$ is defined by
\begin{equation}\label{E:flux}
f(\gamma, \rho) \Def  \gamma \rho \left(1 - \frac{\rho}{\rho_{\max}}\right) = \rho v(\gamma, \rho)
\end{equation}

The mean traffic speed $v:\set{\gamma_\circ, \cdots, \gamma_M} \times [0,\rho_{\max}] \to \RR_+$ is defined by 
\begin{equation}
    \label{E:velocity}
    v(\gamma, \rho) \Def  \gamma \left(1 - \frac{\rho}{\rho_{\max}} \right).
\end{equation}
It should be noted that in Equation \eqref{E:velocity},  $v(\gamma, \rho_{\max}) = 0$ and $v(\gamma, 0) = \gamma$, the maximum permitted speed.

The ODEs describe the trajectories of the $N$ controlled vehicles. In particular, the function $t \mapsto y_i(t)$ denotes the trajectory of the $i$-th AV at time $t$. Then, for $i = 1,\cdots, N$ we define the velocity of the controlled vehicle by
\begin{equation}
    w(y_i, \rho) \Def \min \set{V(y_i), v(\gamma(y_i), \rho)}.
\end{equation}
where, $v(\gamma, \rho)$ is defined as in \eqref{E:velocity} and $V(\cdot)$ is the speed of the controlled vehicle such that $V^{(m)} = V(x) \Def \Bu(t_\circ, x)$ for $x \in  I_m$ where the target speed $\Bu(t_\circ, x)$ (RL suggested speed) at the time step $t_\circ$ is  
 the maximal speed of the controlled vehicle at any $x \in \mathcal I$ region and at the current time step.


The inequality in \eqref{E:main} represents the reduction of the flow caused by the controlled vehicles. In particular, for any $i = 1,\cdots, N$
\begin{equation}
\begin{split}
    F_\alpha(y_i, \dot y_i) &\Def \arg \max_{\rho} \left\{ f_\alpha(\gamma(y_i(t)), \rho)- \dot y_i(t) \rho \right\}\\
    & = \frac{\alpha \rho_{\max}}{4 \gamma(y_i(t))} \left(\gamma(y_i(t)) - \dot y_i(t) \right)^2
\end{split}\end{equation}
where, 
\begin{equation}
    \label{E:reduced_flux}
    f_\alpha(\gamma, \rho) \Def \gamma \rho \left(1 - \frac{\rho}{\alpha \rho_{\max}} \right) , \quad \alpha \in (0, 1)
\end{equation}
Here $\alpha$ represents the reduction rate of the capacity caused by the controlled vehicles. 

\begin{definition}[Weak solution]\label{def:weak_sol}
    The n-tuple  $(\rho, y_i) \in  C(\RR_+; L^1_{\loc}(\RR; [0,1])) \times W^{1,1}_{\loc}(\RR_+, \RR)$ for $i=1,\dots, N$ is a solution of Cauchy problem \eqref{E:main} with initial value $\rho(0, x) = \rho_\circ(x)$, if \begin{enumerate}
    \item function $(t, x) \mapsto \rho(t, x)$ satisfies
    \begin{equation*}
        \int_{\RR_+} \int_{\RR } (\rho \partial_t \varphi + f(\gamma, \rho) \partial_x \varphi) dx dt + \int_{\RR}\rho_\circ(x) \varphi(0, x) dx = 0,
    \end{equation*}
    for all $\varphi \in C_c^\infty(\RR_+ \times \RR)$.
    \item the function $y_i$ for $i=1, \dots, N$ is Carath\'eodory solution, i.e. for a.e. $t \in \RR_+$, in \eqref{E:main}
    \begin{equation}\label{E:BN_ODE}
        y_i(t) = y_{\circ,i} + \int_0^t \omega(y_i(s), \rho(s, y_i(s)+)) ds.
    \end{equation}
    In other words, $y_i \in \AC([0, T]; \RR)$ for any $T >0$, where $\AC$ is the class of absolutely continuous functions. 
    \item The bottleneck constraint is satisfied in the sense that, for a.e. $t \in \RR_+$
    \begin{equation}\label{E:BN_cap}
        \lim_{x \to  y_i(t) \pm} f(\gamma(x), \rho(t, x)) - \rho(t, x)\dot y_i(t) \le F_\alpha(y_i(t), \dot y_i(t)). 
    \end{equation}
\end{enumerate}
\end{definition}

\begin{theorem}[Existence of solution]\label{th:existence}
    Let the initial condition $\rho_\circ \in BV(\RR; [0, \rhomax]) \cap L^1(\RR)$.  Then the PDE-ODE problem \eqref{E:main} has a weak solution in the sense of Definition \ref{def:weak_sol}.
\end{theorem}

\begin{proof}[\small Sketch of the Proof for Theorem \ref{th:existence}]
The complete proof can be found in \cite{matin2023existence}. The model \eqref{E:main} admits classical and non-classical solutions due to the presence of the inequality constraints. 
For the general (non-constant) $\gamma$ the flux function $f(\gamma, \rho)$ contains jump discontinuities. The interaction of classical and non-classical waves with such discontinuities creates a resonant system (coinciding eigenvalues) and hence a non-strict hyperbolic system. In that sense, the uniform total variation of the approximate solutions will be lost and therefore the required compactness theorem for showing the convergence of the approximate solutions does not hold. 
The approach adopted in this proof is the wavefront tracking method. We start with introducing the Riemann solution in this case. Let's consider the initial data
\begin{equation}\label{E:Riemann_condition}
\rho(0, x) = \begin{cases}\rho_L & x <0 \\ \rho_R & x>0 \end{cases}, \quad 
    \gamma(x) = \begin{cases} \gamma_L & x <0 \\
    \gamma_R & x>0 \end{cases}.
\end{equation}
The Riemann solution is denoted by a vector-valued function 
\begin{multline*} \mathcal R: \\
(\rho_L, \rho_R; \gamma_L, \gamma_R) \in [0, 1]^2 \times \set{\gamma_\circ, \cdots, \gamma_{r_M}}^2 \to L_{\loc}^1(\RR; [0,1])\end{multline*}
and is defined according to the minimum jump entropy condition; see for example \cite{temple1982global}. 
In what follows, we will denote with the superscript $^{(R)}$ and $^{(L)}$ the quantities corresponding to the right and left hand side data $(\rho_R, \gamma_R)$ $(\rho_L, \gamma_L)$, respectively. 

\begin{figure}[h!]
    \centering 
    \includegraphics[width=3in]{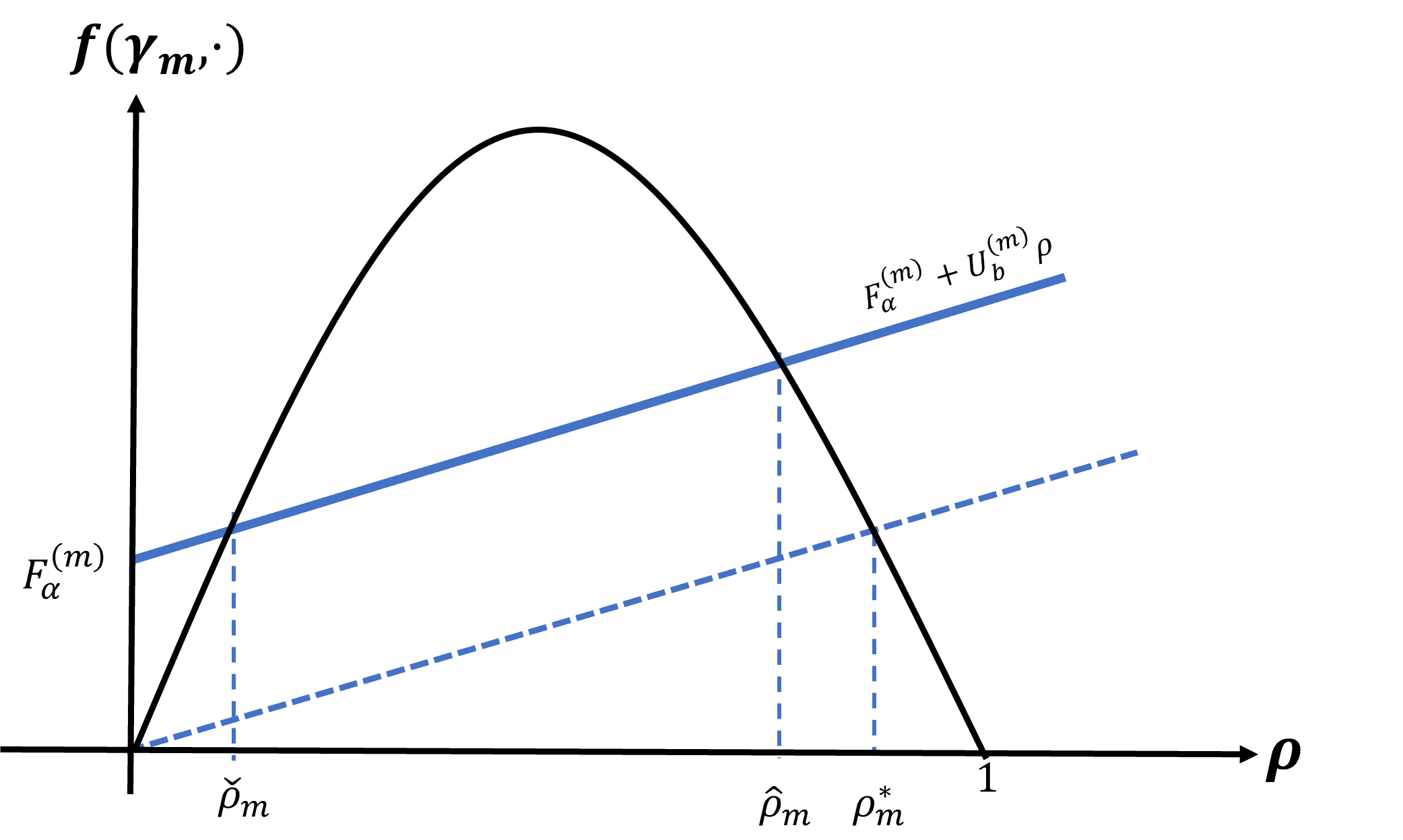}
    \caption{Greenshields Fundamental diagram. The blue solid line represents  the flux constraint in \eqref{E:main} in a certain region $I_m$. For $\rho \in (\check \rho_m, \hat \rho_m)$, i.e. above the solid line, the solution violates the inequality constraint. For $\rho \in [0, \check \rho] \cup [\hat \rho, \rho^*)$ the solution is admissible with $\dot y_i(t) = V^{(m)}$ and for $\rho \in (\rho^*, 1]$ the solution is admissible with $\dot y_i(t) = v(\gamma_m, \rho(t, y_i(t))$.}
    \label{fig:flux_newCoordinates}
\end{figure}
\begin{definition}[Riemann Solution] \label{def:Riemann_sol_new} A Riemann solver 
\begin{equation*} 
\Riemann^\alpha:[0,1]^2 \times \set{\gamma_\circ, \cdots, \gamma_{r_M}}^2 \mapsto L^1(\RR; [0,1])\end{equation*}
for \eqref{E:main} is defined as follows
\begin{enumerate} 
\item If $f(\gamma_R, \Riemann(\rho_L, \rho_R; \gamma_L, \gamma_R)(V^{(R)})) > F_\alpha^{(R)} + V^{(R)} \Riemann(\rho_L, \rho_R; \gamma_L, \gamma_R)(V^{(R)})$, then
\begin{flalign*}
    \Riemann^\alpha (\rho_L, \rho_R; \gamma_L, \gamma_R)(\nicefrac{x}{t}) \Def 
    \begin{cases}
        \Riemann(\rho_L, \hat \rho_R; \gamma_L, \gamma_R)(\nicefrac xt) & \text{for $\frac xt < V^{(R)}$} \\
        \Riemann(\check \rho_R, \rho_R; \gamma_R)(\nicefrac{x}{t}) & \text{for $\frac xt \ge V^{(R)}$}
    \end{cases} 
\end{flalign*}
In this case, we have a non-classical shock between $\hat \rho_R$ and $\check \rho_R$ and we have $y_i(t) = V^{(R)} t$. 

\item If 
\begin{flalign*}
    V^{(R)} \Riemann(\rho_L, \rho_R; \gamma_L, \gamma_R)(V^{(R)})
    \le f(\gamma_R, \Riemann(\rho_L, \rho_R; \gamma_L, \gamma_R)(V^{(R)}))
    \le F_\alpha^{(R)} + V^{(R)} \Riemann(\rho_L, \rho_R; \gamma_L, \gamma_R)(V^{(R)})
\end{flalign*} then 

\begin{flalign*}
    \Riemann^\alpha(\rho_L, \rho_R; \gamma_L, \gamma_R)(\nicefrac{x}{t}) 
    \Def \Riemann(\rho_L, \rho_R; \gamma_L, \gamma_R)(\nicefrac{x}{t}), \quad  y_i(t) = V^{(R)} t.
\end{flalign*}

\item If $f(\gamma_R, \Riemann(\rho_L, \rho_R; \gamma_L, \gamma_R)(V^{(R)}))\le V^{(R)} \Riemann(\rho_L, \rho_R; \gamma_L, \gamma_R)(V^{(R)}) $, then 
\begin{flalign*}
    &\Riemann^\alpha(\rho_L, \rho_R; \gamma_L, \gamma_R)(\nicefrac{x}{t})
    \Def \Riemann(\rho_L, \rho_R; \gamma_L, \gamma_R)(\nicefrac{x}{t}), \quad y_i(t) = v(\gamma_R, \rho_R) t. 
\end{flalign*}

\end{enumerate}
\end{definition}
Figure \ref{fig:flux_newCoordinates} gives an illustrative reference to some of the notation used in the Definition \ref{def:Riemann_sol_new}.
Using the Riemann problem according to the Definition \ref{def:Riemann_sol_new}, we can approximate the initial data $\rho_\circ$ by a piecewise constant, and of bounded variation function $\rho_\circ^{(n)}$ and consequently we can create a sequence of approximate solutions $(\rho^{(n)}, y_i^{(n)})$. As mentioned before, it is known that the $\rho^{(n)}$ is not of bounded variation and the convergence of such sequence to the solution of the Cauchy problem \eqref{E:main} cannot be shown. To show the convergence, for each fixed $\gamma$ we introduce a homeomorphism $(\gamma, \rho) \mapsto \psi(\gamma, z)$. Denoting by $z^{(n)}(t, x) = \psi(\gamma,\rho^{(n)}(t, x))$ and by investigating all possible interactions between the classical and non-classical waves, we can show that
\begin{align*}
    \totvar{z^{(n)}(t, \cdot)}{\RR} \le \totvar{z_{\circ}(\cdot)}{\RR}+ \totvar{\gamma(\cdot)}{\RR} + C,
\end{align*}
Furthermore, using rigorous analysis, we can show that 
\begin{align}
  \sup_{t>0} \norm{z^{(n)}(t, \cdot)}_{L^\infty(\RR)} & \le \frac 14 \max_{m} \gamma_m \le C \label{E:uniform_bound} \\ 
    \norm{z^{(n)}(t, \cdot) - z^{(n)}(s, \cdot)}_{L^1(\RR)} & \le \mathbf C_\ell(t -s), \quad \text{for any $0< s< t$} \label{E:modulus_continuity_z}
\end{align}
Putting all together, we can show by compactness theorem that the sequence $(z^{(n)})_{n \in \NN}$ converges in $L^1_{\loc}(\RR)$ up to a subsequence. Therefore, using the homeomorphism $\psi$, the convergence $(\rho^{(n)})_{n \in \NN}$ in $L^1_{\loc}(\RR)$ can be concluded. 
\end{proof}
\paragraph{\textbf{Evaluating the Performance of the Speed Profile}}
In the evaluation step, two updates are considered. First, the density function $\rho(t, x)$ 
 is used in calculating $\Bv(t_\circ + 1, x)$. Secondly, as defined in MDP, the performance of the policy, i.e., the reward function $r(t_\circ)=R(s(t_\circ),a(t_\circ),s(t_\circ+1))$, is evaluated by the throughput at the bottleneck. The \textit{Actor-Critic} (AC) algorithm plays a pivotal role in \textit{evaluating and updating} the policy for target speed profile generation. In this section, we discuss the structure of the applied AC algorithm and detail the mechanism for updating policy parameters.

The AC algorithm comprises two primary components: the Actor and the Critic. The Actor is responsible for determining the actions based on the current policy, whereas the Critic evaluates these actions by computing the value function.

\begin{itemize}
    \item \textbf{Actor}: Given a TSE $\Bv(t_\circ, x)$, the Actor outputs a distribution over possible desired density according to a policy denoted by $\pi_{\theta}(\rho_b|\Bv(t_\circ, x))$, where $\theta$ is the policy parameter.
    \item \textbf{Critic}: The Critic estimates a value function denoted by $A_w(\Bv(t_\circ, x))$ of the current state, where $w$ is the value function parameter. This function represents the expected cumulative bottleneck throughput from state $\Bv(t_\circ, x)$ onward.
\end{itemize}

The actor is updated by performing gradient ascent on the expected cumulative reward. The parameter update procedure is shown in the Algorithm \ref{algo:ac}.
\begin{algorithm}[h] 
\caption{RL Policy Training Procedure}
\begin{algorithmic}[1]
\State Initialize Actor parameters $\theta$ and Critic parameter $w$
\State Initialize a speed profile $\Bv(t_\circ,x )$ at a fixed time step $t_\circ$. 
\While{not terminated}
    \State Choose desire density $\rho_b$ using policy $\pi_\theta$
    \State Obtain target speed profile and implement to controlled vehicles.
    \State Obtain the bottleneck throughput as $r$ and the TSE of next step $\Bv(t_\circ+1, x)$ by solving the Equation (\ref{E:main})
    \State Compute the Critic's TD error: 
    \begin{equation*}
    \delta = r + \gamma A_w(\Bv(t_\circ+ 1, x) - A_w(\Bv(t_\circ, x))
    \end{equation*}
    \State Update the Actor's parameters:
    \begin{equation*}
    \theta \leftarrow \theta + \alpha \delta \nabla_\theta \log \pi_\theta(\rho_b|\Bv(t_\circ, x))
    \end{equation*}
    \State Update the Critic's parameters:
    \begin{equation*}
    w \leftarrow w + \beta \delta \nabla_w A_w(\Bv(t_\circ, x))
    \end{equation*}
    \State $\Bv(t_\circ, x) \leftarrow \Bv(t_\circ+1, x)$
\EndWhile
\end{algorithmic}
\label{algo:ac}
\end{algorithm}

\begin{sidebarauthor}{Optimal Planner}{Arwa AlAnqary}

As an alternative to the RL-based planner, we propose a planner based on a \textit{model predictive control }(MPC) framework to provide an optimization-based benchmark. This framework consists of two main steps: the prediction step, which estimates the arrival time based on the enhanced TSE data provided by the VSL framework, followed by a planning step, which gives an optimal response to the estimated traffic condition while respecting physical constraints. 


\paragraph{\textbf{Estimated time of arrival (ETA)}} 
 
We consider the spatial horizon $[s_0,\ s_0 + L]$ for $L > 0$ with a not necessarily equidistant grid $\mathcal{S} = [s_0, s_1, s_2, \dots s_l]$, where $s_0 < s_1 < \dots < s_l = s_0 + L$. 
The grid point correspond to the pre-designed divisions of the TSE data. 
With initial time $t_0$, we construct an arrive time estimator which estimates the time $t_i$ it would take a vehicle to arrive to point $s_i$. 
This estimator utilizes the speed information provided by the enhanced TSE and uses the assumption of constant traffic state in each segment $[s_i, s_{i+1}]$. 
Let $\hat{v}(s)$ be the velocity profile as a function of position, then the ETA $t_i$ is computed as
\begin{equation}
    \hat{t}_i = \hat{t}_{i-1} + \int_{s_{i-1}}^{s_i} v(s) \text{d}s
\end{equation}
This results in the set of estimated arrival times $\mathcal{T} = [\hat{t}_0, \hat{t}_1, \hat{t}_2, \dots, \hat{t}_l ]$. 
We can then interpolate the spatial grid $\mathcal{S}$ and the corresponding arrival times $\mathcal{T}$ to construct a trajectory $\hat{s}(t)$. 

\paragraph{\textbf{Speed Planning}} The planning step finds a target trajectory based on the estimated trajectory $\hat{s}(t)$. 
Let us define the quantities of the target trajectory $s^p(t)$, $v^p(t)$, and $a^p(t)$ as the target position, speed, and acceleration respectively. These quantities are governed by the dynamics described by the state-space system in \eqref{eq:ss}.  
\begin{equation}
\label{eq:ss}
    \begin{bmatrix}
        \dot{x}^{p}(t) \\
        \dot{v}^{p}(t)
    \end{bmatrix}
    = \begin{bmatrix}
        0 & 1\\
        0 & 0 
    \end{bmatrix} 
    \begin{bmatrix}
        x^{p}(t) \\
        v^{p}(t)
    \end{bmatrix} + 
    \begin{bmatrix}
        0\\1
    \end{bmatrix} a^{p}(t)
\end{equation}

The task of regulating traffic as well as some physical constraints impose certain conditions that needs to be satisfied by the the planned trajectory.
First we require the speed trajectory to fall within certain speed limits $v(t) \in [v_{\min}, v_{\max}]$ for all time $t$. 
Secondly, we require the target trajectory to have a later arrival time than the estimated one, that is for all $t$ the target trajectory must satisfy $s(t) \leq \hat{s}(t)$. 
This aims to ensure that the target trajectory does not produce unrealistically large velocities that can not be realized given the current state of traffic. 
Finally, the target trajectory should not cause large delays in the arrival time compared to the estimated ones, that is $\hat{s}(t) - s(t) \leq d_{\max}$ where $d_{\max}$ is the maximum delay parameter. 
Among all target trajectories that satisfy these conditions, we want to select the smoothest trajectory as measured by the $\ell^{2}$-norm of the acceleration trajectory $a^{p}(t)$. 
We translate the above into the following optimal control problem over the horizon $t \in [t_0, t_l]$. 

\begin{align}
    \min_{s^{p}, v^{p}, a^{p}} & \quad 
    \int_{t_0}^{t_l} (a^{p}(t))^{2} \\
    \text{s.t.} & \quad     \begin{bmatrix}
        \dot{x}^{p}(t) \\
        \dot{v}^{p}(t)
    \end{bmatrix}
    = \begin{bmatrix}
        0 & 1\\
        0 & 0 
    \end{bmatrix} 
    \begin{bmatrix}
        x^{p}(t) \\
        v^{p}(t)
    \end{bmatrix} + 
    \begin{bmatrix}
        0\\1
    \end{bmatrix} a^{p}(t) \\
    & \quad v_{\min} \leq v^{p}(t) \leq v_{\max}  \\
    & \quad \hat{s}(t) - s(t) \leq d_{\max} \\
    & \quad \hat{s}(t) - s(t) \geq 0 
\end{align}


In order to solve the above optimal control problem efficiently, we introduce a time discretization of the problem over the grid $\mathcal{T}^p = [t_0, t_1, \dots t_m]$. The discretization grid $\mathcal{T}^p$ is a refinement of the arrival times grid $\mathcal{T}$. This discretization results in a linearly constrained quadratic program that can be solved efficiently using generic quadratic optimization solvers. 

\label{sbar:optimal_planner}

\end{sidebarauthor}

\subsection{Vehicle Controllers} 
Another important component of the proposed hierarchical VSL framework is the microscopic traffic control algorithms deployed in individual vehicles. 
In the most general sense, these controllers receive the target speed information as well as locally sensed input from the vehicles surroundings and produce an acceleration control signal to drive the vehicle. Jang et al. \cite{JangReinforcement} details the methodology of the RL-based vehicle controller tested in MVT.
%

The controller in  \cite{JangReinforcement} will be used to produce the simulation results in the "\nameref{sec:simulation}" section. This controller was also used in the road experiments reported in the "\nameref{sec:field_test_analysis}" section.

\section{Numerical Experiments} 
\label{sec:simulation}
We design an array of experiments to numerically evaluate the performance of the proposed VSL framework. We also investigate the robustness of this framework to common limitations in traffic state data that are encountered in practice such as latency and noise. 

\subsection{Simulator}  
In order to test our controller, we built a mixed-autonomy numerical simulator that represents the road portion of the I-24 highway shown in Figure \ref{fig:I-24 network}, see \cite{lichtle2022deploying} for details on the design of the simulator. The simulator was designed for a single lane scenario where two types of vehicles were modeled: human-driven vehicles whose longitudinal behavior is governed by a car-following model and automated vehicles whose behavior is governed by the controller described above. For modeling human behavior, we use the well-known \textit{Intelligent Driver Model} (IDM) \cite{treiber2000congested}. 

\paragraph{\textbf{Test environment}} 
The proposed VSL framework is evaluated on a baseline scenario which comprised of a standing bottleneck in the context of highway merging.  The test environment consisted of a 600-meter-long single lane weaving area where we implemented a forced speed limit to emulate the characteristics of a bottleneck. 
The simulation worked as follows. 
At each time step, we calculate the instantaneous density $\rho$ within the bottleneck area using the number of vehicles $n_{veh}$ and the length of the bottleneck area $L_{BN}$. Then, we derived the corresponding bottleneck speed $V_{BN}$  using a fundamental diagram, calibrated using field data from I-24. 


\paragraph{\textbf{Evaluation metrics}}
The effectiveness of the proposed method was evaluated based on several metrics to provide a comprehensive assessment of system performance.

\textbf{Throughput}. We calculated the throughput at various positions along the road, specifically downstream, in the middle, and upstream of the bottleneck area. Throughput, which is the number of vehicles passing through a point or segment of the road per unit of time, is a direct measure of traffic flow efficiency.

\textbf{\textit{Miles Per Gallon} (MPG)}. System MPG is another key metric for assessing the fuel efficiency of the vehicles within the simulation. A higher MPG indicates better fuel economy and, by extension, improved environmental sustainability.

\textbf{\textit{Vehicle Miles Traveled} (VMT)}. We also considered the system VMT, a measure of the total distance traveled by all vehicles in the system over the simulation time. 

\textbf{System Average Speed}. The system average speed was calculated to evaluate the general speed at which vehicles were able to travel within the domain of the simulation.

\textbf{Speed Standard Deviation}. We also assessed the standard deviation of the speed, providing a measure of the variability in vehicle speeds. A lower standard deviation indicates more uniform speeds and smoother traffic flow.

\paragraph{\textbf{Performance Analysis}}

\begin{figure*}[htbp]
    \centering
    \includegraphics[width=\linewidth]{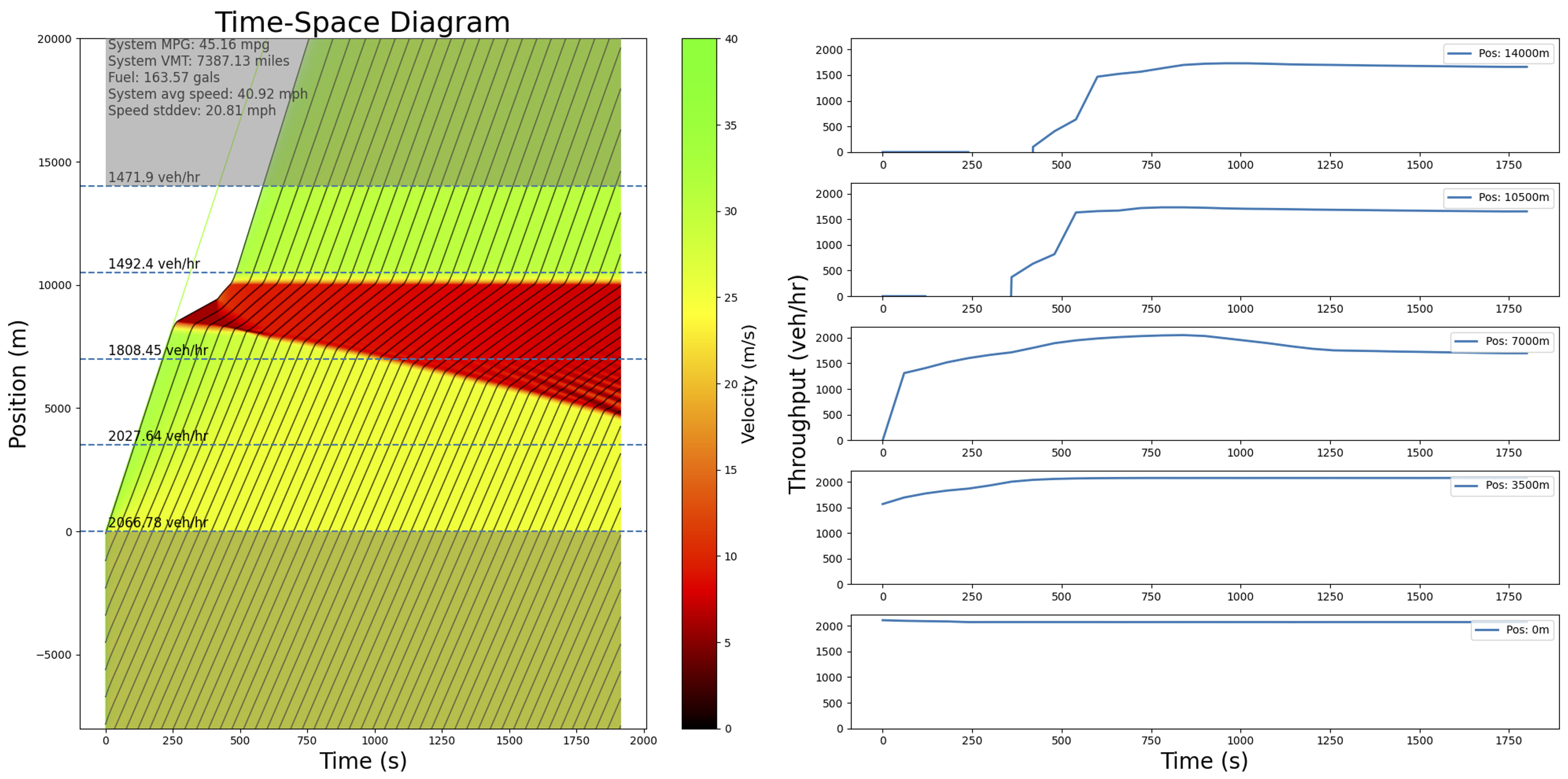}
    \caption{\textbf{IDM baseline scenario}. In black AV trajectories.  Bottleneck throughput: 1492.4 veh/hr, System MPG: 45.16 mpg, System VMT: 7387.13 miles, Fuel consumption: 163.57 gals, System average speed: 40.92 mph, Speed standard deviation: 20.81 mph.}
    \label{fig:sim_baseline}
\end{figure*}

\begin{figure*}[htbp]
    \centering
    \includegraphics[width=\linewidth]{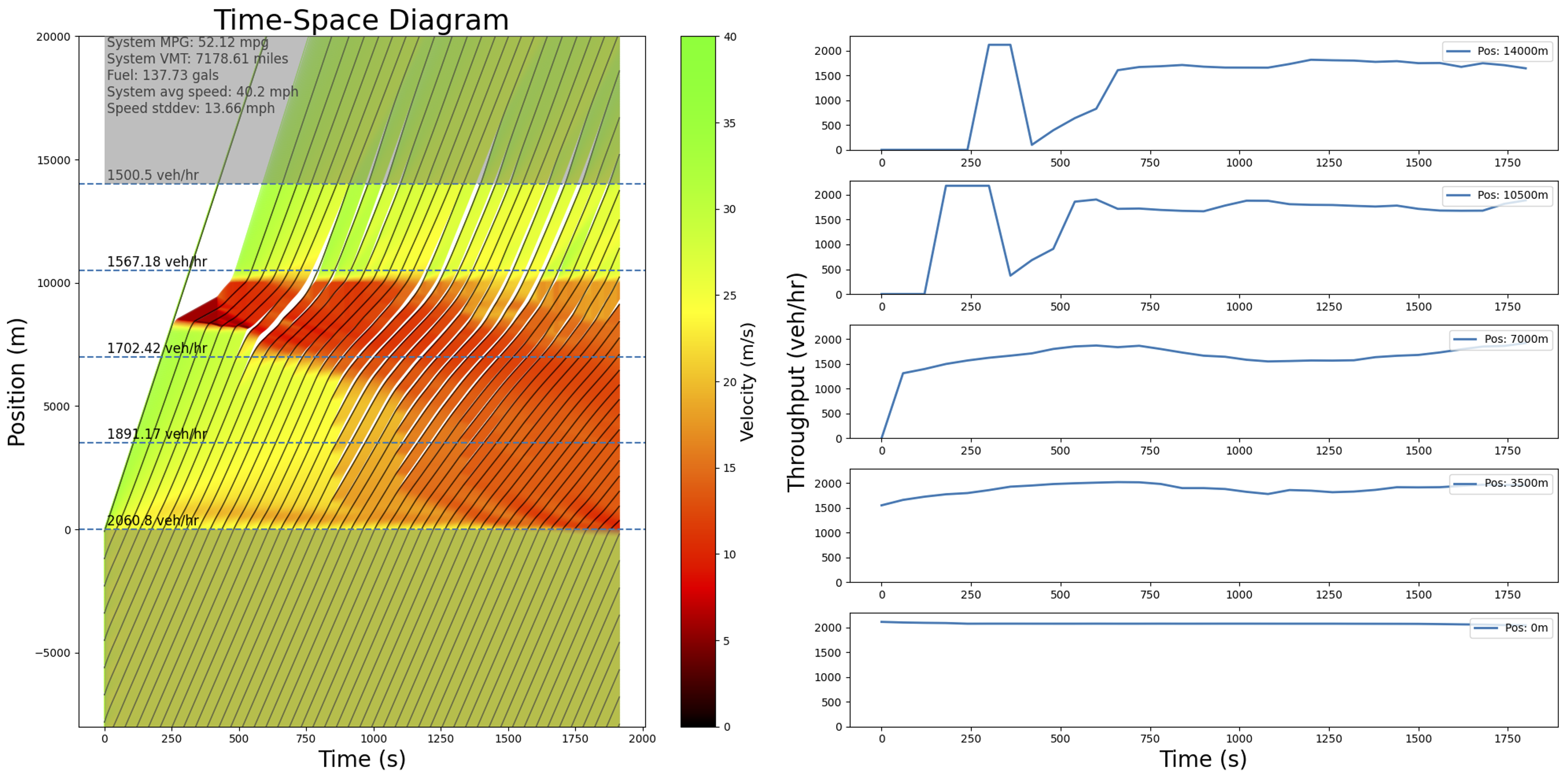}
    \caption{\textbf{Speed Planner scenario}. AV trajectories are shown in black curves. Bottleneck throughput: 1567.18 veh/hr (+5.01\%), System MPG: 52.12 mpg (+15.41\%), System VMT: 7178.61 miles (-2.82\%), Fuel consumption: 137.73 gals (-34.14\%), System average speed: 40.2 mph (-1.76\%), Speed standard deviation: 13.66 mph (-34.36\%).}
    \label{fig:sim_sp}
\end{figure*}

The effectiveness of the Speed Planner is evident by comparing the simulation results of the IDM baseline scenario and the Speed Planner controlled scenario with the 4\% AV penetration rate, as shown in Figures \ref{fig:sim_baseline} and \ref{fig:sim_sp}.

In the \textbf{IDM baseline scenario} (Figure \ref{fig:sim_baseline}), the system exhibits a bottleneck throughput of 1492.4 veh/hr, a system MPG of 45.16 mpg, and a total VMT (Vehicle Miles Traveled) of 7387.13 miles. The fuel consumption stands at 163.57 gallons, with the system average speed being 40.92 mph and a speed standard deviation of 20.81 mph. The \textbf{Speed Planner scenario} (Figure \ref{fig:sim_sp}) demonstrates significant improvements in several key metrics. The bottleneck throughput has increased by 5.01\% to 1567.18 veh/hr, and the system MPG has seen a substantial rise of 15.41\% to 52.12 mpg. Fuel consumption has been reduced by 34.14\% to 137.73 gallons, showcasing the eco-friendliness and cost-effectiveness of the Speed Planner. Most notably, the speed standard deviation has been reduced by a significant 34.36\% to 13.66 mph, indicating a more uniform and stable traffic flow when the Speed Planner is in control. However, as the trade-off for achieving better fuel efficiency and throughput by homogenizing traffic flow, the system average speed has slightly decreased by 1.76\% to 40.2 mph, and the system VMT has decreased by 2.82\% to 7178.61 miles. 

\section{Field Test Analysis}
\label{sec:field_test_analysis}
This section aims to provide a detailed examination of the MegaVanderTest, the operational traffic experiment with the largest deployment of AVs, and its implications using a dataset~\cite{LeeMegacontrollerCSM} created using the I-24 MOTION system~\cite{gloudemans202324}.

\subsection{MegaVanderTest}
\begin{figure*}[h!]
    \centering
    \includegraphics[width=\linewidth]{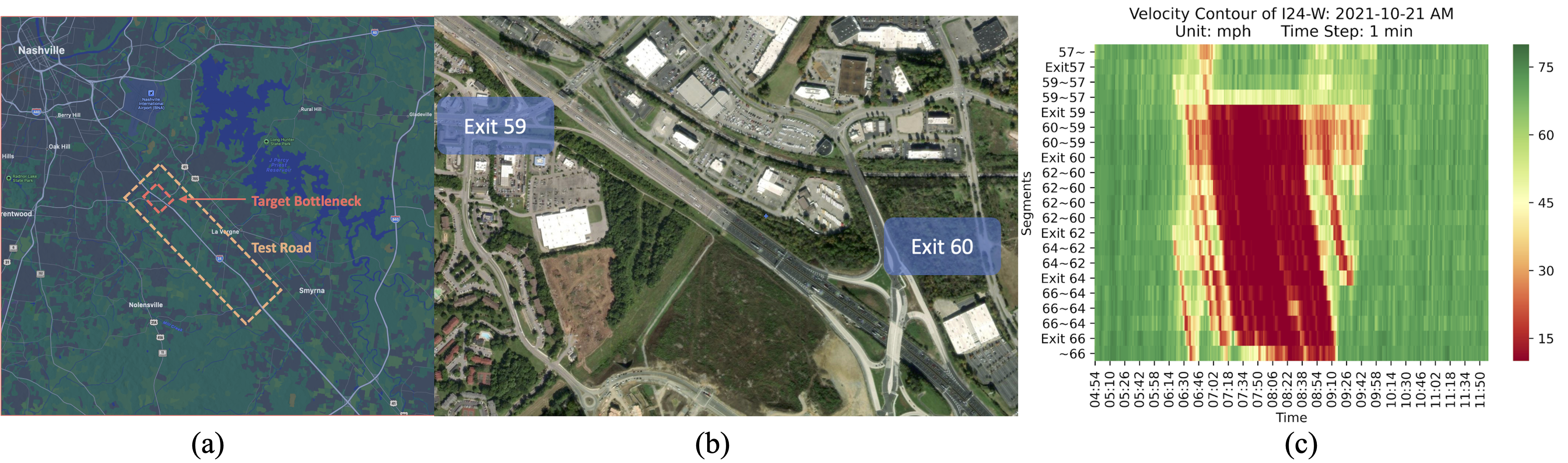}
    \caption{MVT test road: (a) An illustration of the targeted highway network within this study (I-24 Westbound in Nashville, Tennessee), seen within the dashed line region. (b) Satellite view of the bottleneck target area between Exit 60 and Exit 59, TN. (c) Velocity heatmap of morning peak hour during a classical work day, Oct 21, 2021. Source: INRIX~\cite{cookson2017inrix}.}
    \label{fig:I-24 network}
\end{figure*}

In mid-November 2022, the CIRCLES Consortium~\cite{CIRCLES} carried out a large-scale traffic experiment on a 14.5-km stretch of Interstate I-24 in Nashville, Tennessee, depicted in Figure~\ref{fig:I-24 network}. Over recent years, this network has gathered attention from researchers due to the creation of the I-24 MOTION testbed~\cite{gloudemans202324,gloudemans2020interstate}. Efforts have been made to reconstruct its features, understand the implemented variable speed limit control system~\cite{9922471,10207650}, and tackle driving characteristics that lead to energy inefficiencies~\cite{lichtle2022deploying, kardous2022rigorous, hayat2022holistic, lee2021integrated}. 

The large-scale traffic experiment, which we name "MegaVanderTest", deployed a fleet of 100 vehicles consisting of three market-selling car models including Nissan Rogue, Toyota RAV4, and Cadillac XT5 onto the studied network during the morning commute over the course of five days (11/14/2022 - 11/18/2022). Each vehicle was equipped with a variety of control algorithms, which we name "MegaController" (details are available in paper \cite{LeeMegacontroller}), overriding or modifying the cruise control system, and designed to automatically adjust the longitudinal speed of the vehicle to improve the overall flow of traffic---turning each car into its own ``robot traffic manager" ~\cite{fu2023massive}.

To achieve this large-scale operation, more than 50 CIRCLES \cite{CIRCLES} researchers from around the world gathered into a field headquarters in a converted office space in Antioch, Tennessee. Each morning of the experiment, trained drivers drove the control vehicles on the designed routes through the I-24 MOTION testbed~\cite{gloudemans202324}. For the interested reader we refer to \cite{AmeliLivetraffic} for more details about the organization of the experiment. As the drivers traversed their routes, researchers collected traffic data from both the vehicles and the I-24 MOTION traffic monitoring system.



The hierarchical framework and the Speed Planner introduced in "\nameref{s:methodology}", as part of the "MegaController", was deployed and tested in MVT. The following section presents the operational analysis based on the I24 Motion dataset.


\begin{figure*}[htbp]
    \centering
    \includegraphics[width=\linewidth]{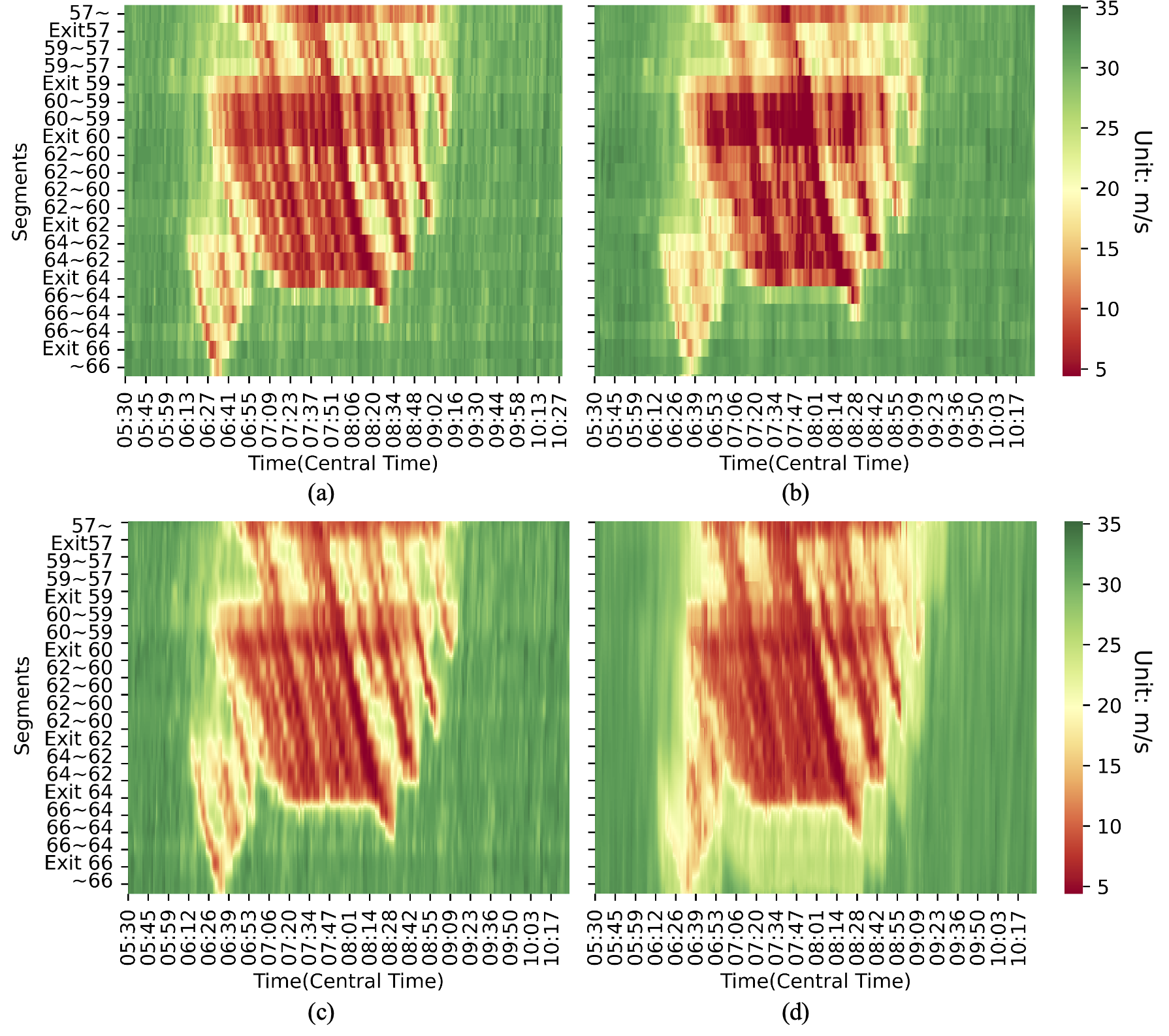}
    \caption{Comprehensive visualization of the Speed Planner data processing stages on November 18th, 2022. \textbf{(a) INRIX}: The heatmap of the raw INRIX data, illustrating the initial TSE. \textbf{(b) Prediction}: The output heatmap of the prediction module, aiming to forecast the traffic condition 3 minutes ahead to counteract INRIX's inherent latency. We can see that the prediction captured the location of the standing bottleneck around Exit 59 and kept accurate tracking of shockwaves. \textbf{(c) Fusion}: The output heatmap of the fusion module, where real-time vehicle observations are integrated with the predicted traffic data, resulting in a high-resolution lane-level (TSE). \textbf{(d) Target}: The heatmap of the final target speed profile derived from the processed data, guiding vehicle controllers for optimal navigation. A distinct early deceleration zone (buffer area) can be observed upstream of the congestion as the yellow transition zone in the heatmap.}
    \label{fig:inrix_2_target}
\end{figure*}

\begin{figure*}[htbp]
    \centering
    \includegraphics[width=14cm]{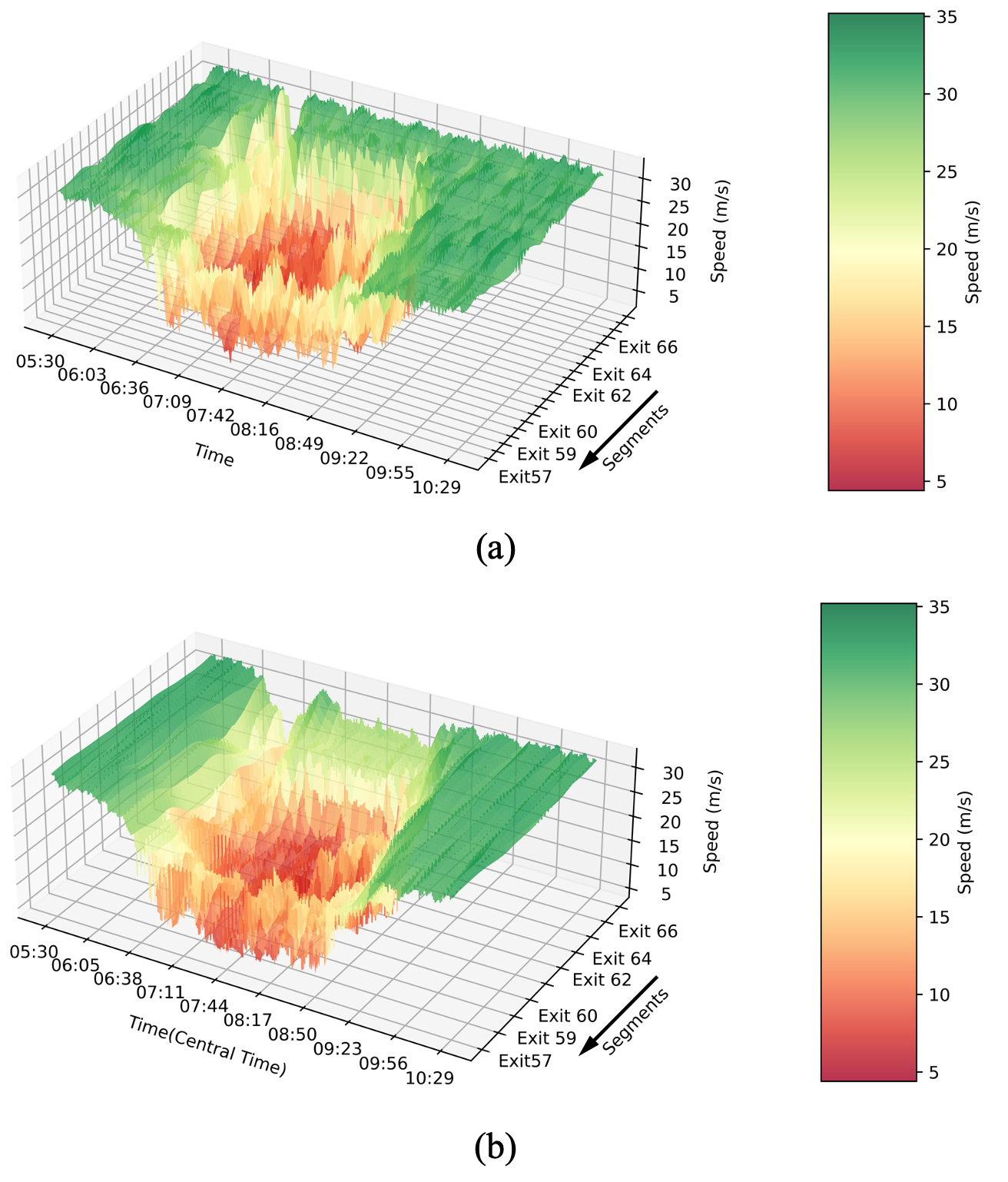}
    \caption{Speed surfaces of \textbf{(a) Enhanced TSE} and \textbf{(b) Target Speed}. In (b) the overall speed surface is smoother. Upstream of the congestion (Exit 62 -- Exit 66 in the figure), our system designs a smooth speed gradient for vehicles about to enter the queue. The goal is to guide the traffic to slow down in advance. (Arrow indicates the traffic flow direction.)
    }
    \label{fig:3d_inrix}
\end{figure*}

\begin{figure*}[h!]
    \centering
    \includegraphics[width=0.8\linewidth]{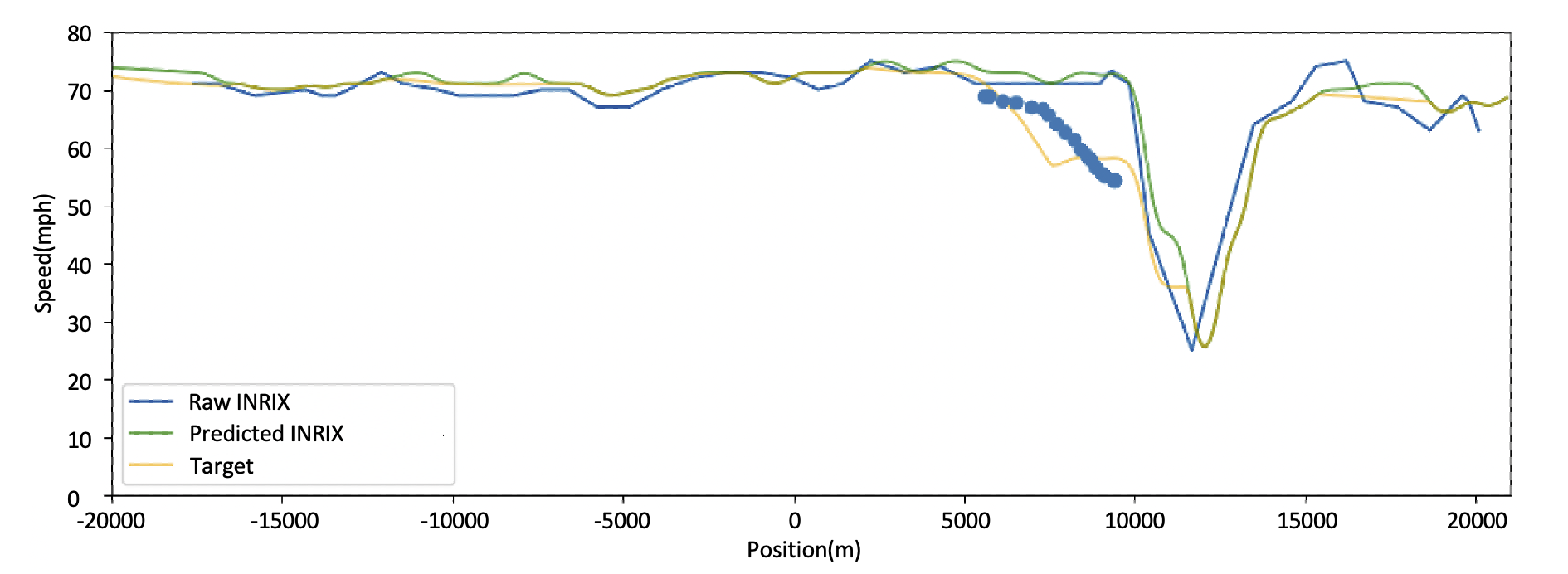}
    \caption{Speed profile illustration: Raw INRIX, Predicted INRIX, Target Speed Profile. Blue scatters indicate the ping of a single AV during that target update interval.}
    \label{fig:av_traj}
\end{figure*}

\begin{figure*}[htbp]
    \centering
    \includegraphics[width=0.925\linewidth]{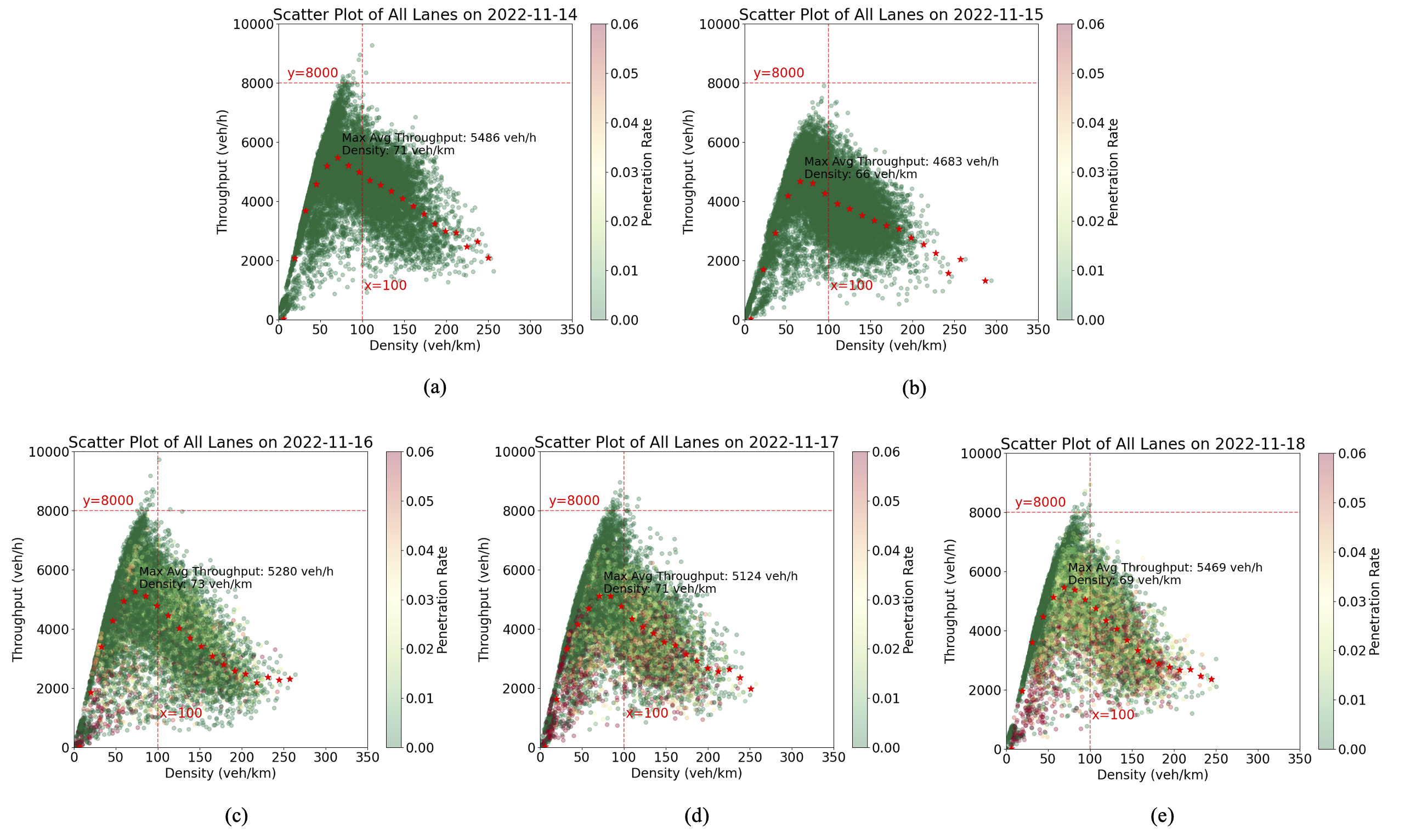}
    \caption{Scatters of density-throughput colored by the penetration rate of controlled vehicles with Speed Planner engaged. Red stars are the mean throughput for measurement points binned by density (x-axis). In the order of days of the MVT week: \textbf{(a) Monday}: 80+ Nissans running stock ACC without connection to Speed Planner, which is considered as the human baseline. \textbf{(b) Tuesday}: No controlled vehicle on the road due to inclement weather. \textbf{(c) Wednesday}: $~80+$ Nissans deployed with Speed Planner Engaged on the westbound, stock ACC elsewhere. \textbf{(d) Thursday}: $~80+$ Nissans deployed with Speed Planner Engaged. High server latency was observed. \textbf{(e) Friday}: 97 Nissans deployed with Speed Planner Engaged. Jang et al. \cite{JangReinforcement} provides a detailed deployment timeline on the vehicle side.}
    \label{fig:fd_scatters}
\end{figure*}

\begin{figure*}[htbp]
    \centering
    \includegraphics[width=\linewidth]{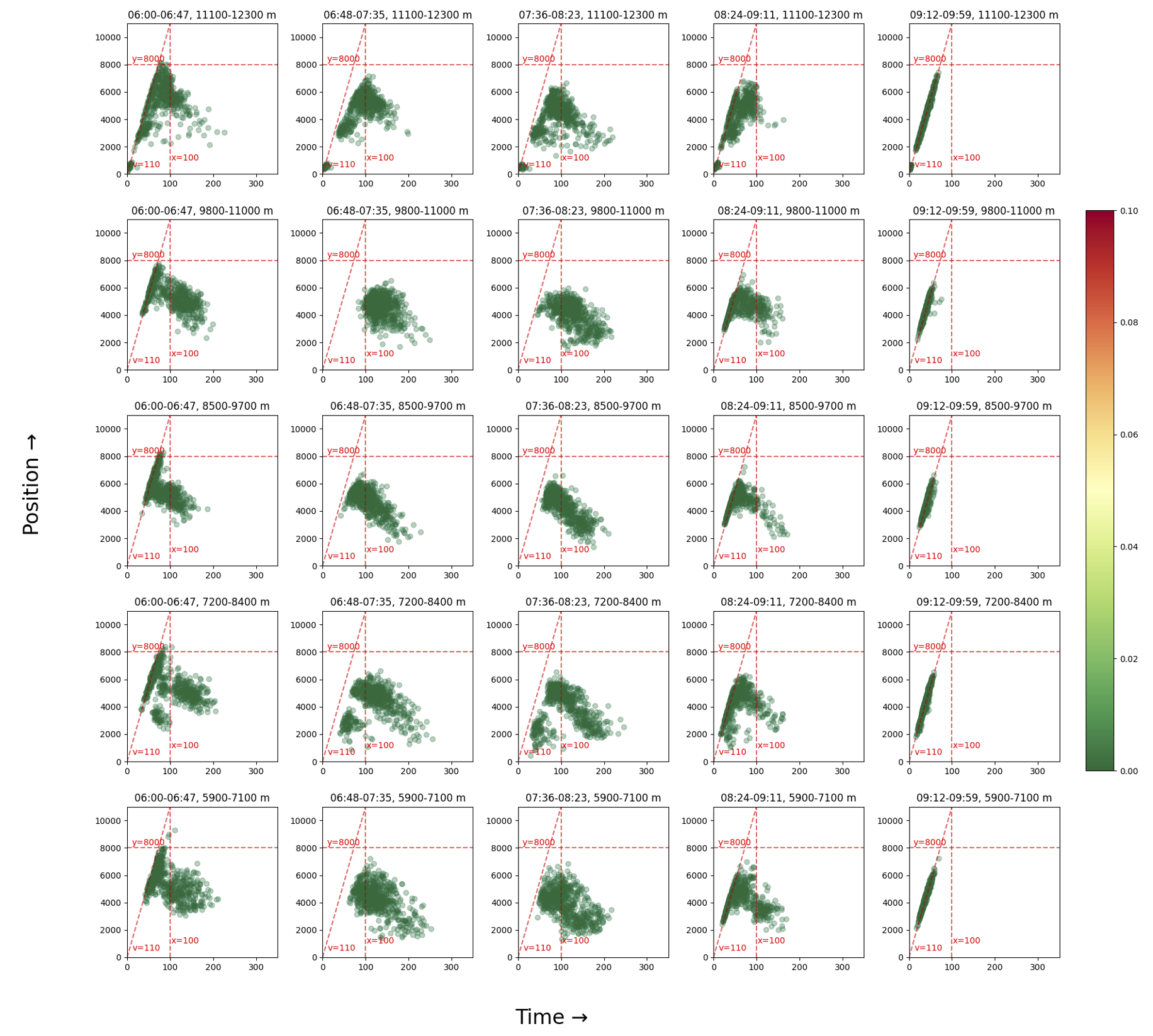}
    \caption{Density-throughput scatters grid of \textbf{Monday} divided by time-space windows. Colored by the penetration rate of controlled vehicles with Speed Planner engaged.}
    \label{fig:mon_grid}
\end{figure*}

\begin{figure*}[htbp]
    \centering
    \includegraphics[width=\linewidth]{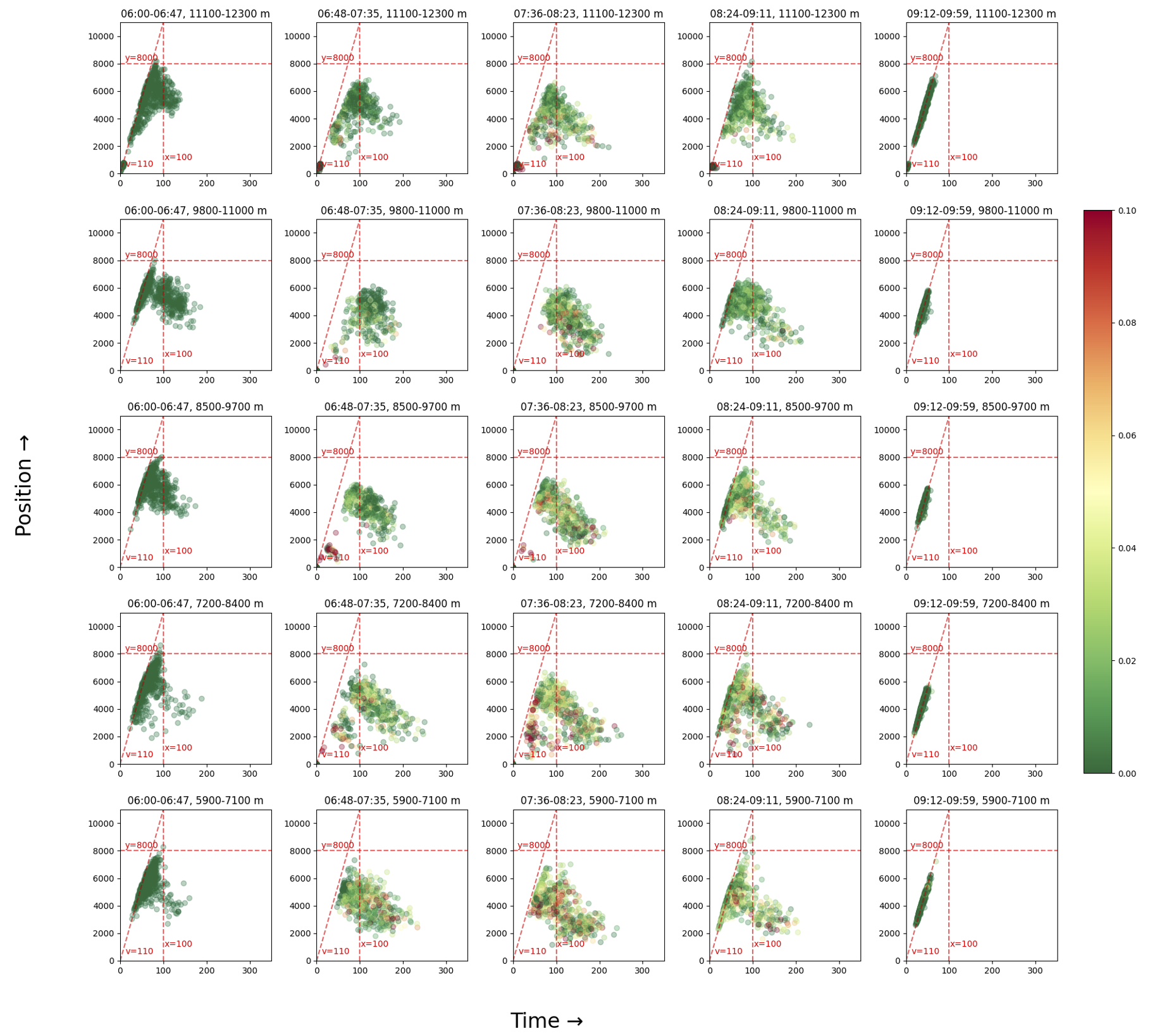}
    \caption{Density-throughput scatters grid of \textbf{Friday} divided by time-space windows. Colored by the penetration rate of controlled vehicles with Speed Planner engaged. Noteworthy density and throughput variations exist in the four scatters in the range 5900-8400m, 07:36-09:11.}
    \label{fig:fri_grid}
\end{figure*}

\begin{figure*}[htbp]
    \centering
    \includegraphics[width=\linewidth]{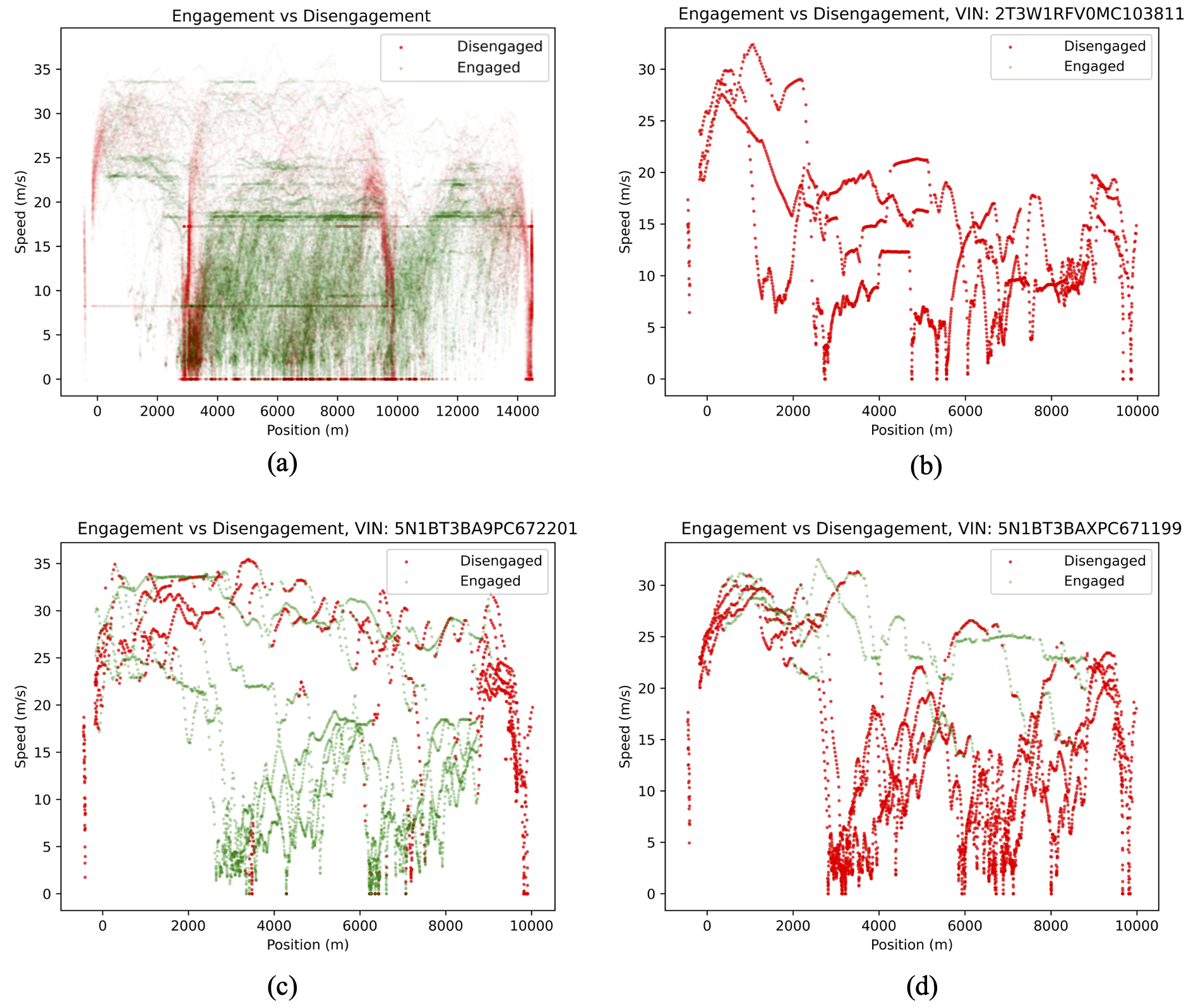}
    \caption{Scatters reflecting controller engagement pattern on November 18th, 2022: (a) Overall controller engagement trend. Mostly, drivers followed the instructions and activated the controller on the westbound, especially in the congested sections. Four vertical red lines are the terminal ramps for two designed routes, where drivers take over the control to exit and re-enter the highway. However, the individual difference is noteworthy: (b) An individual driver who did not activate the controller during the whole experiment. (c) An individual driver activated the controller during the congestion period and took over the control in the free-flow period. (d) An individual driver had the opposite preference from the driver in (c). The controller was only activated in some free-flow periods.}
    \label{fig:engagement}
\end{figure*}

\begin{figure*}[htbp]
    \centering
    \includegraphics[width=\linewidth]{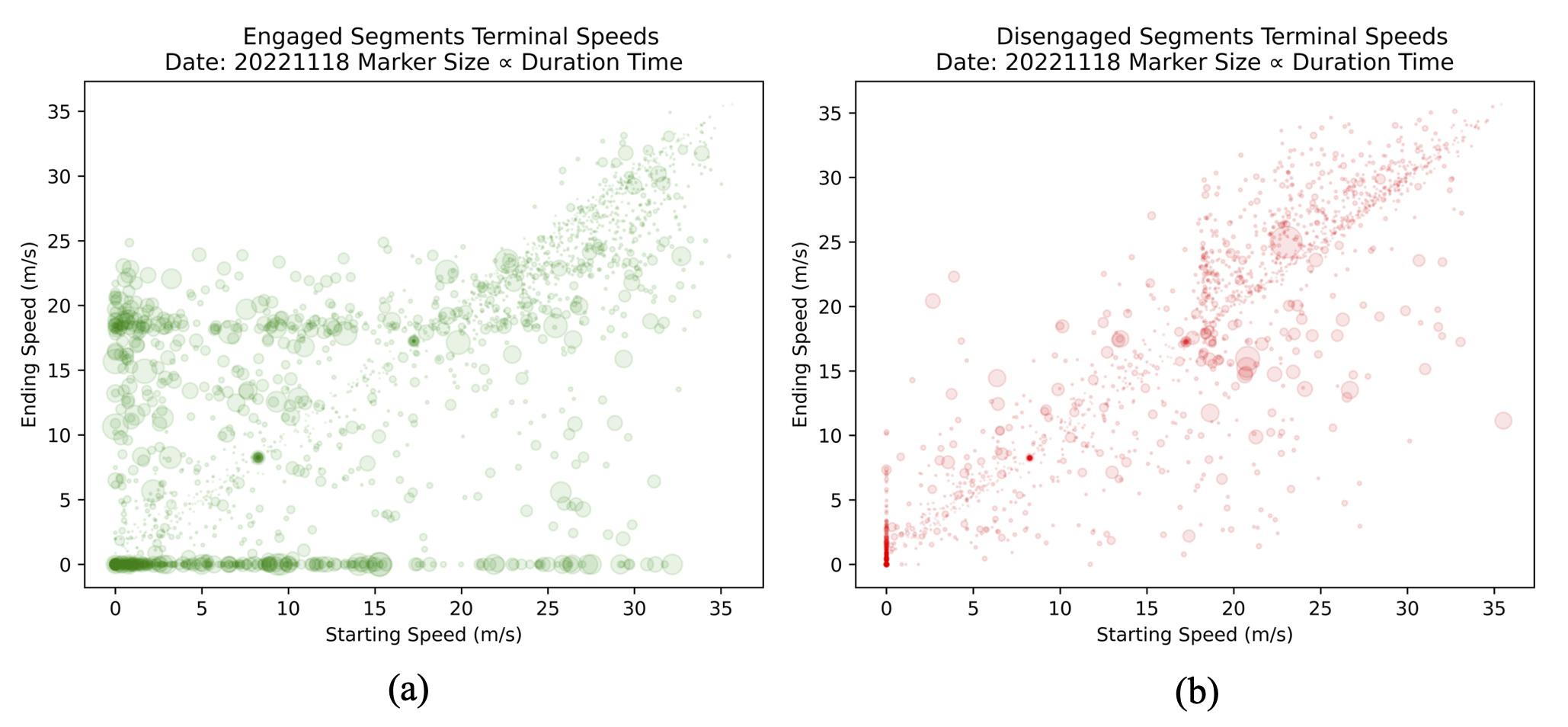}
    \caption{Scatters of the starting speed-ending speed when driver: \textbf{(a) engaged} the controller. The significant clustering at the level where the ending speed equals 0 suggests that drivers tend to take over control when restarting from a full stop during congestion. The aggregation around the 18m/s level may be caused by the driver taking over as the vehicle reaches the end of the route and pulls onto the ramp; when driver \textbf{(b) disengaged} the controller, which indicates similar driving pattern as (a).} 
    \label{fig:eng_terminal}
\end{figure*}

\begin{figure*}[htbp]
    \centering
    \includegraphics[width=\linewidth]{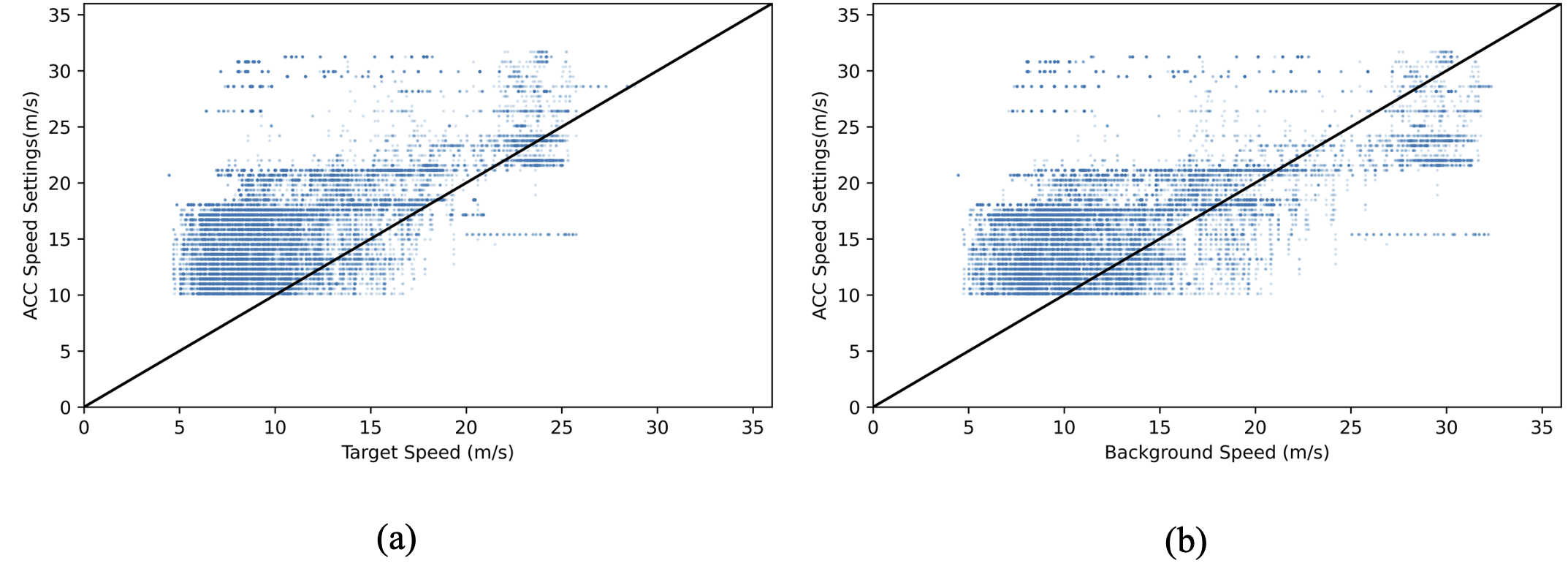}
    \caption{Scatter of ACC speed setting output by vehicle controller to \textbf{(a) Target speed} assigned by Speed Planner. Correlation coef.: 0.5604. \textbf{(b) Traffic Background speed} - INRIX. Correlation coef.: 0.5659.} 
    \label{fig:acc_target}
\end{figure*}

\subsection{Performance Analysis}

The performance analysis of the proposed system in MVT is based on I-24 MOTION \cite{gloudemans202324,gloudemans2020interstate}, a state-of-the-art measurement system to collect and analyze vehicle trajectory data on a section of the I-24 highway in Tennessee. The I-24 MOTION system extracts vehicle trajectory datasets from the video images captured by  276 high-resolution traffic cameras. These datasets contain the position of each vehicle on the highway, as well as other supplementary information such as vehicle dimensions and class. The trajectory data is generated using computer vision techniques~\cite{gloudemans2023so,gloudemans2023interstate,gloudemans2021vehicle}, which analyze the video frames to track the movement of individual vehicles over time. The resulting dataset is then stitched~\cite{wang2023onlinemcf},  regularized~\cite{wang2022automatic}, and visualized~\cite{10.1145/3576914.3587710} to create a detailed record of vehicle behavior on the freeway, including speed, acceleration, lane changes, and other maneuvers. Trajectories data are stored in JSON format along with additional metadata such as scene homography, trajectory extraction algorithm settings, and descriptions of data attributes. 


The performance of the Speed Planner can be understood through a series of visualizations and analyses that depict its data processing stages, the impact of its deployment, and the behavior of drivers when interacting with the system.

\textbf{Data Processing Stages}.
Figure \ref{fig:inrix_2_target} provides a comprehensive visualization of the Speed Planner data processing stages on November 18th, 2022. The raw INRIX data, as shown in \textbf{(a)}, serves as the initial TSE. The prediction module, depicted in \textbf{(b)}, forecasts the traffic condition 3 minutes ahead to counteract INRIX inherent latency. We can see that the prediction captured the location of the standing bottleneck around Exit 59 and maintained accurate tracking of shockwaves. The fusion module, illustrated in \textbf{(c)}, integrates real-time vehicle observations with the predicted traffic data, resulting in a high-resolution lane-level TSE. Finally, \textbf{(d)} showcases the heatmap of the final target speed profile derived from the processed data, guiding vehicle controllers for optimal navigation. A distinct early deceleration zone (buffer area) can be observed upstream of the congestion, evident as the yellow transition zone in the heatmap.

Figure \ref{fig:3d_inrix} presents the speed surfaces of both the Enhanced TSE and the Target Speed. The overall speed surface appears smoother, especially at the upstream of congestion (Exit 62 - Exit 66), where our system designs a gentle slope with a smooth speed gradient for vehicles about to enter congestion, guiding traffic to decelerate in advance. In Figure \ref{fig:av_traj}, we present a comparison of these speed profiles: the raw INRIX data, the predicted INRIX output from our prediction module, and the final target speed profile generated by the Speed Planner. A buffer area together with the deceleration area can be observed on the target profile (5000m - 9000m). The blue scatters overlaying the profiles represent pings from an individual AV sent during the target update interval. These pings provide insights into the real-time behavior of the AVs in relation to the recommended speed profiles. The vehicle exhibits a propensity to adhere to the prescribed target speed profile. The divergence between target speed and vehicle ping indicates that the tracking is not enforced due to the necessity of accommodating local microscopic traffic conditions.

\textbf{Deployment Impact}.
Figure \ref{fig:fd_scatters} illustrates the density-throughput scatters of the MVT weekdays, colored by the penetration rate of controlled vehicles with the Speed Planner engaged. 
The density and throughput are calculated using Edie's definition \cite{edie1963discussion} with a $100m \times 60sec$ window. This choice creates a balance between granularity and the ability to capture meaningful traffic patterns obtained by experiment. 
The variations in the scatter plots across different days provide insights into the system behavior under different conditions. Figures \ref{fig:mon_grid} and \ref{fig:fri_grid} further divide it into sub-scatter for Monday (stock ACCs) and Friday (with controlled vehicles), respectively. These grids, divided by time-space windows and colored by the penetration rate of controlled vehicles with the Speed Planner engaged, showcase the system impact at different times and locations. On observing the scatter plots, it is evident that congestion often began before our vehicles entered the roadway. This early onset of congestion led to near-saturated traffic conditions, making it  challenging to produce significant improvements. Such conditions underscore the complexities inherent in operational traffic scenarios and highlight the need for future experiments to consider these dynamics.


\textbf{Driver Engagement Analysis}. In operational field tests, the performance of the overall system is intricately linked to driver behavior, especially in the context of controlled vehicles. Figure \ref{fig:engagement} reflects the controller engagement pattern on November 18th, 2022. The patterns indicate varied levels of driver trust and willingness to activate the controller. While many drivers chose to activate the controller on the westbound, individual differences in engagement patterns are evident. Some drivers chose not to activate the controller throughout the experiment, while others displayed varied preferences based on traffic conditions.

The frequency and manner in which drivers engage or disengage the controller can significantly influence the system performance. Figure \ref{fig:eng_terminal} provides insights into the starting and ending speeds when drivers either engaged or disengaged the controller. In Figure \ref{fig:eng_terminal}(a), which represents the speeds when the controller was engaged, there is a noticeable clustering at the level where the ending speed equals 0. This suggests that drivers tend to take over control, especially when restarting from a full stop during congestion. Additionally, there is an aggregation around the 18m/s level, which may indicate instances where the driver takes over as the vehicle approaches the end of the route and transitions onto the ramp. Figure \ref{fig:eng_terminal}(b), shows the speeds when the controller was disengaged, and it mirrors a similar driving pattern as observed in (a). This consistency in patterns across both engagement and disengagement actions underscores specific scenarios in which drivers tend to take over control, providing insights for future system refinement and driver training.

Furthermore, even when the controller is activated, the system performance can be influenced by the degree to which vehicle controllers adhere to the assigned target speed. Figure \ref{fig:acc_target} presents a scatter plot comparing the ACC speed setting output by the vehicle controller to both the Target speed assigned by the Speed Planner and the Background speed from INRIX. From the correlation coefficients provided in Figure \ref{fig:acc_target}, it's evident that the ACC speed setting has a slightly higher correlation with the Background speed (0.5659) than with the Target speed (0.5604). This suggests that while the vehicle controllers are influenced by the Speed Planner  recommendations, they tend to align more closely with the prevailing traffic conditions as represented by INRIX. This indicates that the vehicle controller tested in MVT might prioritize harmonizing with the immediate traffic environment to ensure safety and smooth driving experiences. A challenge emerges when our vehicles, following the Speed Planner recommendations, drive at speeds significantly different from surrounding traffic. Even if the recommended speeds are optimized for overall traffic flow, driving much slower or faster than nearby vehicles might lead to social acceptance issues. Such discrepancies can make drivers feel out of sync with the general traffic, potentially leading to discomfort or reduced trust in the system.

\section{Conclusion}

The development and deployment of the Speed Planner in the MVT system represent a significant step forward in the realm of traffic management and vehicle control. Through a hierarchical framework, the system effectively integrates macro TSE from external sources like INRIX with micro observations from probe vehicles, offering a comprehensive view of the traffic environment.

Our performance analysis, as detailed in the preceding sections, underscores the system capability to process and predict traffic data, fuses it with real-time vehicle observations, and derives optimal target speed profiles. The visualizations provide a clear representation of the data processing stages, from raw INRIX data to the final target speed profile. The system design ensures smoother speed gradients, especially in congestion-prone areas, aiming to enhance overall traffic flow.

However, as the Driver Engagement Analysis highlighted, the  effectiveness of such systems is closely tied to driver behavior. The extent to which drivers trust and engage with the controller, as well as their adherence to the recommended speeds, can significantly influence the system performance. The potential social acceptance challenges, especially when our vehicles drive at speeds divergent from surrounding traffic, underline the importance of considering human factors in the design and deployment of automated systems.

In conclusion, while the Speed Planner showcases promising capabilities in optimizing traffic flow and vehicle control, its full potential can only be realized with a holistic approach that considers both technological advancements and human-centric factors. Future work should focus on enhancing system robustness, improving driver trust and engagement, and addressing potential social acceptance issues to ensure seamless integration of such systems into our daily commutes.

\section{Acknowledgement}

This material is based upon work supported by the National Science Foundation under Grants CNS-1837244 (A.~Bayen), CNS-1837652 (D.~Work), CNS-1837481 (B.~Piccoli), CNS-1837210 (G.~Pappas), CNS-1446715 (B.~Piccoli), CNS-1446690 (B.~Seibold), CNS-1446435 (J.~Sprinkle, R.~Lysecky), CNS-1446702 (D.~Work), CNS-2135579 (D.~Work, A.~Bayen, J.~Sprinkle, J.~Lee), and by the French CNRS under the grant IEA SHYSTRA and PEPS JCJC (A.~Hayat). This material is based upon work supported by the U.S.\ Department of Energy’s Office of Energy Efficiency and Renewable Energy (EERE) under the Vehicle Technologies Office award number CID DE--EE0008872. The views expressed herein do not necessarily represent the views of the U.S.\ Department of Energy or the United States Government.  The authors are grateful for the additional support provided by C3AI, Amazon AWS, Siemens, Toyota, GM, Nissan, Caltrans, CCTA, and the Tennessee Department of Transportation.

\newpage
\section{Author Biography}

\begin{IEEEbiography}{{H}an Wang}{\,}(hanw@berkeley.edu) As of the submission of this thesis, Han is working on his Ph.D. in Transportation Engineering, as well as an M.S. degree in the Department of Electrical Engineering and Computer Sciences at U.C. Berkeley. His research focuses on the decentralized control of massive AV platforms and driving assistance utilizing computer vision techniques.
\end{IEEEbiography}

\begin{IEEEbiography}{{Z}he Fu}{\,}(zhefu@berkeley.edu) is a Ph.D. candidate in Transportation Engineering and a M.S. candidate at the Department of Electrical Engineering and Computer Sciences at UC Berkeley. She has research affiliations with the Institute of Transportation Studies (ITS), Berkeley Artificial Intelligence Research (BAIR), California Partners for Advanced Transportation Technology (California PATH), and Berkeley Deep Drive (BDD). Her research interests lie in traffic control and machine learning in mixed-autonomy environments.
\end{IEEEbiography}

\begin{IEEEbiography}{{J}onathan W.~Lee}{\,}(jonny5@berkeley.edu) received the B.S.~degree in engineering physics from the University of California, Berkeley and M.S.~and Ph.D.~degrees in mechanical engineering from Rice University. At Sandia National Laboratories (2011--13), he completed his postdoctoral appointment studying electrical and material properties via molecular dynamics simulations. He subsequently served as a senior data scientist and product manager on various teams at Uber Technologies, Inc.~(2014-2019). Since 2019, he has served as an engineering manager at the University of California, Berkeley and the program manager and Chief Engineer of CIRCLES.
\end{IEEEbiography}

\begin{IEEEbiography}{Arwa Alanqary}{\,} (arwa@berkeley.edu)
Arwa Alanqary received her B.S. in mechanical engineering from Alfaisal University in 2017 and M.S. degree in computational science and engineering from Massachusetts Instutte of Technology in 2019. She is currently a
 Ph.D. student in electrical engineering and computer sciences at the University of California, Berkeley. 
Her research interests include optimal control and optimization and learning in multi-agent systems.   
\end{IEEEbiography}

\begin{IEEEbiography}{{H}ossein Nick Zinat Matin}{\,}(h-matin@berkeley.edu) Hossein is a postdoctoral researcher at the University of California Berkeley. His research interest is partial differential equations and control and their applications in the theory of traffic flow. 
\end{IEEEbiography}

\begin{IEEEbiography}{{D}aniel Urieli}{\,}(daniel.urieli@gm.com) received his Ph.D.~in artificial intelligence  (AI) from the University of Texas at Austin, where he developed AI agents that won the Power Trading Agent Competition, as well as the simulated robot soccer league in the international RoboCup competition. Daniel is a Staff Researcher at General Motors, where he has co-founded the autonomous driving research group and led research on applying AI for congestion reduction and electric vehicle energy management.
\end{IEEEbiography}

\begin{IEEEbiography}{{S}haron Hornstein}{\,}(sharon.hornstein@gm.com) is a researcher at the General Motors R\&D site in Israel. Sharon received her B.Sc., M.Sc. and Ph.D. degrees in Mechanical Engineering from the Technion, Israel. Her research interest is focused on the dynamic modeling of non-linear systems.   
\end{IEEEbiography}

\begin{IEEEbiography}{{A}bdul Rahman Kreidieh}{\,}(aboudy@berkeley.edu) is Ph.D. student at UC Berkeley, Department of Civil and Environmental Engineering.
\end{IEEEbiography}

\begin{IEEEbiography}{{R}aphael Chekroun}{\,}(raphael.chekroun@minesparis.psl.eu) is a research scientist at Valeo in France and a PhD candidate in the Center for Robotics, Mines Paris, PSL University. He worked as a visiting scholar at UC Berkeley with the CIRCLES consortium. His research is focused on AI for autonomous driving and its application to traffic flow regulation.
\end{IEEEbiography}

\begin{IEEEbiography}{William Barbour}{\,}(william.w.barbour@vanderbilt.edu) is research scientist at Vanderbilt University and the Institute for Software Integrated Systems. He is a co-principle investigator on the I-24 MOTION testbed and works on other large transportation systems projects across the state of Tennessee. His technical focus is on intelligent transportation systems, artificial intelligence, and data science.
\end{IEEEbiography}

\begin{IEEEbiography}{{W}illiam A. Richardson}{\,}(william.a.richardson@vanderbilt.edu) is graduate student researcher at Vanderbilt University.
\end{IEEEbiography}

\begin{IEEEbiography}{Dan Work}{\,}(dan.work@vanderbilt.edu) is a Chancellor Faculty Fellow and professor in civil and environmental engineering, computer science, and the Institute for Software Integrated Systems at Vanderbilt University. He has held research appointments at the University of Illinois at Urbana-Champaign (2010–17), Institute for Pure and Applied Mathematics (2015, 2020), Microsoft Research Redmond (2009), and Nokia Research Center Palo Alto (2007–09). His research interests are broadly in transportation cyber-physical systems, with many applications in highway traffic sensing and control.
\end{IEEEbiography}

\begin{IEEEbiography}{{B}enedetto Piccoli}{\,}(piccoli@camden.rutgers.edu) is University Professor at Rutgers University-Camden. He also served as Vice Chancellor for Research. He received his Ph.D. degree in applied mathematics from the Scuola Internazionale Superiore di Studi Avanzati (SISSA), Trieste, Italy, in 1994. He was a Researcher with SISSA from 1994 to 1998, an Associate Professor with the University of Salerno from 1998 to 2001, and a Research Director with the Istituto per le Applicazioni del Calcolo “Mauro Picone” of the Italian Consiglio Nazionale delle Ricerche (IAC-CNR), Rome, Italy, from 2001 to 2009. Since 2009, he has been the Joseph and Loretta Lopez Chair Professor of Mathematics with the Department of Mathematical Sciences, Rutgers University–Camden, Camden, NJ, USA.
\end{IEEEbiography}

\begin{IEEEbiography}{Benjamin Seibold}{\,}(seibold@temple.edu) 
is a Professor of Mathematics and Physics, and the Director of the Center for Computational Mathematics and Modeling, at Temple University. His research areas, funded by NSF, DOE, DAC, USACE, USDA, and PDA, are computational mathematics (high-order methods for differential equations, CFD, molecular dynamics) and applied mathematics and modeling (traffic flow, invasive species, many-agent systems, radiative transfer).
\end{IEEEbiography}

\begin{IEEEbiography}{Jonathan Sprinkle}{\,}(jonathan.sprinkle@vanderbilt.edu)
Jonathan Sprinkle is a Professor of Computer Science at Vanderbilt University since 2021. Prior to joining Vanderbilt he was the Litton Industries John M. Leonis Distinguished Associate Professor of Electrical and Computer Engineering at the University of Arizona, and the Interim Director of the Transportation Research Institute. From 2017-2019 he served as a Program Director in Cyber-Physical Systems and Smart \& Connected Communities at the National Science Foundation in the CISE Directorate. 
\end{IEEEbiography}

\begin{IEEEbiography}{{A}lexandre M. Bayen}{\,}(bayen@berkeley.edu)
is the Associate Provost for Moffett Field Program Development at UC Berkeley, and the Liao-Cho Professor of Engineering at UC Berkeley. He is a Professor of Electrical Engineering and Computer Science and of Civil and Environmental Engineering (courtesy). He is a Visiting Professor at Google. He is also a Faculty Scientist in Mechanical Engineering, at the Lawrence Berkeley National Laboratory (LBNL). From 2014 - 2021, he served as the Director of the Institute of Transportation Studies at UC Berkeley (ITS). )
\end{IEEEbiography}

\begin{IEEEbiography}{Maria\textcolor{white}{a}Laura\textcolor{white}{a}Delle\textcolor{white}{a}Monache}{\,}(mldellemonache@berkeley.edu) is an assistant professor in the Department of Civil and Environmental Engineering at the University of California, Berkeley. Dr. Delle Monache’s research lies at the intersection of transportation engineering, mathematics, and control and focuses on modeling and control of mixed autonomy large-scale traffic systems.
\end{IEEEbiography}

\bibliographystyle{IEEEtran}
\bibliography{CIRCLES-key-papers, reference}

\begin{thebibliography}{10}
\providecommand{\url}[1]{#1}
\csname url@samestyle\endcsname
\providecommand{\newblock}{\relax}
\providecommand{\bibinfo}[2]{#2}
\providecommand{\BIBentrySTDinterwordspacing}{\spaceskip=0pt\relax}
\providecommand{\BIBentryALTinterwordstretchfactor}{4}
\providecommand{\BIBentryALTinterwordspacing}{\spaceskip=\fontdimen2\font plus
\BIBentryALTinterwordstretchfactor\fontdimen3\font minus \fontdimen4\font\relax}
\providecommand{\BIBforeignlanguage}[2]{{%
\expandafter\ifx\csname l@#1\endcsname\relax
\typeout{** WARNING: IEEEtran.bst: No hyphenation pattern has been}%
\typeout{** loaded for the language `#1'. Using the pattern for}%
\typeout{** the default language instead.}%
\else
\language=\csname l@#1\endcsname
\fi
#2}}
\providecommand{\BIBdecl}{\relax}
\BIBdecl

\bibitem{seo2017traffic}
T.~Seo, A.~M. Bayen, T.~Kusakabe, and Y.~Asakura, ``Traffic state estimation on highway: A comprehensive survey,'' \emph{Annual reviews in control}, vol.~43, pp. 128--151, 2017.

\bibitem{papageorgiou2008effects}
M.~Papageorgiou, E.~Kosmatopoulos, and I.~Papamichail, ``Effects of variable speed limits on motorway traffic flow,'' \emph{Transportation Research Record}, vol. 2047, no.~1, pp. 37--48, 2008.

\bibitem{ADFFP98}
A.~Alessandri, A.~Di~Febbraro, A.~Ferrara, and E.~Punta, ``Optimal control of freeways via speed signalling and ramp metering,'' \emph{Control Engineering Practice}, vol.~6, pp. 771--780, 1998.

\bibitem{HDSH05}
A.~Hegyi, B.~De~Schutter, and J.~Hellendoorn, ``Optimal coordination of variable speed limit to suppress shock waves,'' \emph{IEEE Transactions on intelligent transportation systems}, vol.~6, no.~1, pp. 102--112, 2005.

\bibitem{HXZ07}
Z.~Hou, J.-X. Xu, and H.~Zhong, ``Freeway traffic control using iterative learning control-based ramp metering and speed signaling,'' \emph{IEEE Transactions on vehicular technology}, vol.~56, no.~2, pp. 466--477, 2007.

\bibitem{HHSSV08}
A.~Hegyi, S.~P. Hoogendoorn, M.~Schreuder, H.~Stoelhorst, and F.~Viti, ``{SPECIALIST}: {A} dynamic speed limit control algorithm based on shock wave theory,'' in \emph{Proceedings of the 11th International IEEE Conference on Intelligent Transportation Systems}, 2008, pp. 827 -- 832.

\bibitem{HHSS09}
A.~Hegyi, S.~P. Hoogendoorn, M.~Schreuder, and H.~Stoelhorst, ``The expected effectivity of the dynamic speed limit algorithm {SPECIALIST} - a field data evaluation method,'' in \emph{Proceedings of the European Control Conference}, 2009, pp. 1770--1775.

\bibitem{HH10}
A.~Hegyi and S.~P. Hoogendoorn, ``Dynamic speed limit control to resolve shock waves on freeways - {F}ield test results of the {SPECIALIST} algorithm,'' in \emph{13th International {IEEE} Annual conference on Intelligent Transportation Systems}, 2010, pp. 519--524.

\bibitem{CPPM10}
R.~C. Carlson, I.~Papamichail, M.~Papageorgiou, and A.~Messmer, ``Optimal motorway traffic flow control involving variable speed limits and ramp metering,'' \emph{Transportation Science}, vol.~44, pp. 238--253, 2010.

\bibitem{CDW11}
C.~Canudas De~Wit, ``Best-effort highway traffic congestion control via variable speed limits,'' in \emph{50th IEEE Conference on Decision and Control and European Control Conference}, 2011.

\bibitem{FC12}
J.~R. Dom\'{i}ngeuz~Frejo and E.~F. Camacho, ``Global versus local {MPC algorithms in freeway traffic control with ramp metering and variable speed limits},'' \emph{IEEE Transactions on intelligent transportation systems}, vol.~13, no.~4, pp. 1556--1565, 2012.

\bibitem{YLLZ13}
X.~Yang, Y.~Lin, Y.~Lu, and N.~Zou, ``Optimal variable speed limit control for real-time freeway congestions,'' in \emph{13th {COTA} International Conference of Transportation Professionals ({CICTP} 2013)}, P.~Social and B.~Sciences, Eds., vol.~96, 2013, pp. 2362--2372.

\bibitem{CVH13}
A.~Csik\'{o}s, I.~Varga, and K.~Hangos, ``Freeway shockwave control using ramp metering and variable speed limits,'' in \emph{21st Mediterranean Conference on Control \& Automation}, 2013, pp. 1569--1574.

\bibitem{delle2017traffic}
M.~L. Delle~Monache, B.~Piccoli, and F.~Rossi, ``Traffic regulation via controlled speed limit,'' \emph{SIAM Journal on Control and Optimization}, vol.~55, no.~5, pp. 2936--2958, 2017.

\bibitem{GGK14}
P.~Goatin, S.~G\"{o}ttlich, and O.~Kolb, ``Speed limit and ramp meter control for traffic flow networks,'' \emph{Engineering Optimization}, vol.~48, no.~7, pp. 1121--1144, 2016.

\bibitem{Rama1999}
P.~Rämä, ``Effects of weather-controlled variable speed limits and warning signs on driver behavior,'' \emph{Transportation research record}, vol. 1678, pp. 34--40, 1999.

\bibitem{hegyi2005}
A.~Hegyi, B.~De~Schutter, and H.~Hellendoorn, ``Model predictive control for optimal coordination of ramp metering and variable speed limits,'' \emph{Transportation research part C: emerging technologies}, vol.~13, no.~3, pp. 181--201, 2005.

\bibitem{han2017}
Y.~Han, D.~Chen, and S.~Ahn, ``Variable speed limit control at fixed freeway bottlenecks using connected vehicles,'' \emph{Transportation Research Part B: Methodological}, vol.~98, pp. 113--134, 2017.

\bibitem{azin202280}
B.~Azin, X.~Yang, M.-P. Consortium \emph{et~al.}, ``I-80 hybrid regulatory speed limit signing design and vsl system evaluation [mpc-22-488],'' Mountain-Plains Consortium, Tech. Rep., 2022.

\bibitem{shukla2017hierarchical}
S.~Shukla and L.~Mili, ``Hierarchical decentralized control for enhanced rotor angle and voltage stability of large-scale power systems,'' \emph{IEEE Transactions on Power Systems}, vol.~32, pp. 4783--4793, 2017.

\bibitem{wang2015decentralized}
Y.~Wang, P.~Yemula, and A.~Bose, ``Decentralized communication and control systems for power system operation,'' \emph{IEEE Transactions on Smart Grid}, vol.~6, pp. 885--893, 2015.

\bibitem{li2021module}
D.~Li and C.~N. Ho, ``A module-based plug-n-play dc microgrid with fully decentralized control for ieee empower a billion lives competition,'' \emph{IEEE Transactions on Power Electronics}, vol.~36, pp. 1764--1776, 2021.

\bibitem{li2017designing}
X.~Li, Q.~Xu, and C.~Chen, ``Designing a hierarchical decentralized system for distributing large-scale, cross-sector, and multipollutant control accountabilities,'' \emph{IEEE Systems Journal}, vol.~11, pp. 2774--2783, 2017.

\bibitem{gessing1985two}
R.~Gessing, ``Two-level hierarchical control for stochastic optimal resource allocation†,'' \emph{International Journal of Control}, vol.~41, pp. 161--175, 1985.

\bibitem{wang2021enhanced}
Y.~Wang, P.~Liu, D.~Liu, F.~Deng, and Z.~Chen, ``Enhanced hierarchical control framework of microgrids with efficiency improvement and thermal management,'' \emph{IEEE Transactions on Energy Conversion}, vol.~36, pp. 11--22, 2021.

\bibitem{stern2018dissipation}
R.~E. Stern, S.~Cui, M.~L. Delle~Monache, R.~Bhadani, M.~Bunting, M.~Churchill, N.~Hamilton, H.~Pohlmann, F.~Wu, B.~Piccoli \emph{et~al.}, ``Dissipation of stop-and-go waves via control of autonomous vehicles: Field experiments,'' \emph{Transportation Research Part C: Emerging Technologies}, vol.~89, pp. 205--221, 2018.

\bibitem{delle2019feedback}
M.~L. Delle~Monache, T.~Liard, A.~Rat, R.~Stern, R.~Bhadani, B.~Seibold, J.~Sprinkle, D.~B. Work, and B.~Piccoli, ``Feedback control algorithms for the dissipation of traffic waves with autonomous vehicles,'' \emph{Computational Intelligence and Optimization Methods for Control Engineering}, pp. 275--299, 2019.

\bibitem{Wu2018}
A.~R. Kreidieh, C.~Wu, and A.~M. Bayen, ``Dissipating stop-and-go waves in closed and open networks via deep reinforcement learning,'' in \emph{2018 21st International Conference on Intelligent Transportation Systems (ITSC)}, 2018, pp. 1475--1480.

\bibitem{jode}
M.~L. Delle~Monache and P.~Goatin, ``Scalar conservation laws with moving constraints arising in traffic flow modeling: an existence result,'' \emph{J. Differential Equations}, vol. 257, no.~11, pp. 4015--4029, 2014.

\bibitem{Cicic2018}
M.~{\v C}i{\v c}i{\'c} and K.~H. Johansson, ``Traffic regulation via individually controlled automated vehicles: a cell transmission model approach,'' in \emph{2018 21st International Conference on Intelligent Transportation Systems (ITSC)}, 2018, pp. 766--771.

\bibitem{Cicic2020}
M.~{\v C}i{\v c}i{\'c}, I.~Mikol{\'a}{\v s}ek, and K.~H. Johansson, ``Front tracking transition system model with controlled moving bottlenecks and probabilistic traffic breakdowns,'' \emph{IFAC-PapersOnLine}, vol.~53, no.~2, pp. 14\,990--14\,996, 2020.

\bibitem{giammarino2020traffic}
V.~Giammarino, S.~Baldi, P.~Frasca, and M.~L. Delle~Monache, ``Traffic flow on a ring with a single autonomous vehicle: An interconnected stability perspective,'' \emph{IEEE Transactions on Intelligent Transportation Systems}, vol.~22, no.~8, pp. 4998--5008, 2020.

\bibitem{bayen2022control}
A.~Bayen, M.~L. Delle~Monache, M.~Garavello, P.~Goatin, and B.~Piccoli, \emph{Control problems for conservation laws with traffic applications: modeling, analysis, and numerical methods}.\hskip 1em plus 0.5em minus 0.4em\relax Springer Nature, 2022.

\bibitem{goatin2022interacting}
P.~Goatin, C.~Daini, M.~L. Delle~Monache, and A.~Ferrara, ``Interacting moving bottlenecks in traffic flow,'' \emph{Networks and Heterogeneous Media}, 2022.

\bibitem{daini2022centralized}
C.~Daini, P.~Goatin, M.~L. Delle~Monache, and A.~Ferrara, ``Centralized traffic control via small fleets of connected and automated vehicles,'' in \emph{2022 European Control Conference (ECC)}.\hskip 1em plus 0.5em minus 0.4em\relax IEEE, 2022, pp. 371--376.

\bibitem{Liard20231190}
T.~Liard, R.~Stern, and M.~L.~D. Monache, ``A pde-ode model for traffic control with autonomous vehicles,'' \emph{Networks and Heterogeneous Media}, vol.~18, no.~3, pp. 1190 -- 1206, 2023.

\bibitem{cicic2020coordinating}
M.~{\v C}i{\v c}i{\'c}, L.~Jin, and K.~H. Johansson, ``Coordinating vehicle platoons for highway bottleneck decongestion and throughput improvement,'' 2020.

\bibitem{Cicic2019}
M.~{\v C}i{\v c}i{\'c} and K.~H. Johansson, ``Energy-optimal platoon catch-up in moving bottleneck framework,'' in \emph{2019 18th European Control Conference (ECC)}, 2019, pp. 3674--3679.

\bibitem{cookson2017inrix}
G.~Cookson and B.~Pishue, ``Inrix global traffic scorecard--appendices,'' \emph{INRIX research}, 2017.

\bibitem{bunting2021libpanda}
\BIBentryALTinterwordspacing
M.~Bunting, R.~Bhadani, and J.~Sprinkle, ``Libpanda: A high performance library for vehicle data collection,'' in \emph{Proceedings of the Workshop on Data-Driven and Intelligent Cyber-Physical Systems}, 2021, pp. 32--40. [Online]. Available: \url{https://dl.acm.org/doi/abs/10.1145/3459609.3460529}
\BIBentrySTDinterwordspacing

\bibitem{bhadani2022strym}
\BIBentryALTinterwordspacing
R.~Bhadani, M.~Bunting, M.~Nice, N.~M. Tran, S.~Elmadani, D.~Work, and J.~Sprinkle, ``Strym: A python package for real-time can data logging, analysis and visualization to work with usb-can interface,'' in \emph{2022 2nd Workshop on Data-Driven and Intelligent Cyber-Physical Systems for Smart Cities Workshop (DI-CPS)}.\hskip 1em plus 0.5em minus 0.4em\relax IEEE, 2022, pp. 14--23. [Online]. Available: \url{https://ieeexplore.ieee.org/abstract/document/9805366}
\BIBentrySTDinterwordspacing

\bibitem{elmadani2021can}
\BIBentryALTinterwordspacing
S.~Elmadani, M.~Nice, M.~Bunting, J.~Sprinkle, and R.~Bhadani, ``From can to ros: A monitoring and data recording bridge,'' in \emph{Proceedings of the Workshop on Data-Driven and Intelligent Cyber-Physical Systems}, 2021, pp. 17--21. [Online]. Available: \url{https://dl.acm.org/doi/abs/10.1145/3459609.3460531}
\BIBentrySTDinterwordspacing

\bibitem{nice2023middleware}
\BIBentryALTinterwordspacing
M.~Nice, M.~Bunting, D.~Work, and J.~Sprinkle, ``Middleware for a heterogeneous cav fleet,'' in \emph{Proceedings of Cyber-Physical Systems and Internet of Things Week 2023}, 2023, pp. 86--91. [Online]. Available: \url{https://dl.acm.org/doi/abs/10.1145/3459609.3460531}
\BIBentrySTDinterwordspacing

\bibitem{fu2023cooperative}
Z.~Fu, A.~R. Kreidieh, H.~Wang, J.~W. Lee, M.~L. Delle~Monache, and A.~M. Bayen, ``Cooperative driving for speed harmonization in mixed-traffic environments,'' in \emph{2023 IEEE Intelligent Vehicles Symposium (IV)}.\hskip 1em plus 0.5em minus 0.4em\relax IEEE, 2023, pp. 1--8.

\bibitem{pananurak2009adaptive}
W.~Pananurak, S.~Thanok, and M.~Parnichkun, ``Adaptive cruise control for an intelligent vehicle,'' in \emph{2008 IEEE International Conference on Robotics and Biomimetics}.\hskip 1em plus 0.5em minus 0.4em\relax IEEE, 2009, pp. 1794--1799.

\bibitem{vaswani2017attention}
A.~Vaswani, N.~Shazeer, N.~Parmar, J.~Uszkoreit, L.~Jones, A.~N. Gomez, {\L}.~Kaiser, and I.~Polosukhin, ``Attention is all you need,'' \emph{Advances in neural information processing systems}, vol.~30, 2017.

\bibitem{chen2004systematic}
C.~Chen, A.~Skabardonis, and P.~Varaiya, ``Systematic identification of freeway bottlenecks,'' \emph{Transportation Research Record}, vol. 1867, no.~1, pp. 46--52, 2004.

\bibitem{kingma2014adam}
D.~P. Kingma and J.~Ba, ``Adam: A method for stochastic optimization,'' \emph{arXiv preprint arXiv:1412.6980}, 2014.

\bibitem{asadi2010role}
B.~Asadi, C.~Zhang, and A.~Vahidi, ``The role of traffic flow preview for planning fuel optimal vehicle velocity,'' in \emph{Dynamic systems and control conference}, vol. 44182, 2010, pp. 813--819.

\bibitem{cui2017stabilizing}
S.~Cui, B.~Seibold, R.~Stern, and D.~B. Work, ``Stabilizing traffic flow via a single autonomous vehicle: Possibilities and limitations,'' in \emph{2017 IEEE Intelligent Vehicles Symposium (IV)}.\hskip 1em plus 0.5em minus 0.4em\relax IEEE, 2017, pp. 1336--1341.

\bibitem{kreidieh2022learning}
A.~R. Kreidieh, Z.~Fu, and A.~M. Bayen, ``Learning energy-efficient driving behaviors by imitating experts,'' in \emph{2022 IEEE 25th International Conference on Intelligent Transportation Systems (ITSC)}.\hskip 1em plus 0.5em minus 0.4em\relax IEEE, 2022, pp. 2689--2695.

\bibitem{daganzo1997fundamentals}
C.~F. Daganzo, \emph{Fundamentals of transportation and traffic operations}.\hskip 1em plus 0.5em minus 0.4em\relax Emerald Group Publishing Limited, 1997.

\bibitem{greenberg1959analysis}
H.~Greenberg, ``An analysis of traffic flow,'' \emph{Operations research}, vol.~7, no.~1, pp. 79--85, 1959.

\bibitem{newell1961nonlinear}
G.~F. Newell, ``Nonlinear effects in the dynamics of car following,'' \emph{Operations research}, vol.~9, no.~2, pp. 209--229, 1961.

\bibitem{puterman2014markov}
M.~L. Puterman, \emph{Markov decision processes: discrete stochastic dynamic programming}.\hskip 1em plus 0.5em minus 0.4em\relax John Wiley \& Sons, 2014.

\bibitem{lebacque1998introducing}
J.-P. Lebacque, J.-B. Lesort, and F.~Giorgi, ``Introducing buses into first-order macroscopic traffic flow models,'' \emph{Transportation Research Record}, vol. 1644, no.~1, pp. 70--79, 1998.

\bibitem{leclercq2004moving}
L.~Leclercq, S.~Chanut, and J.-B. Lesort, ``Moving bottlenecks in lighthill-whitham-richards model: A unified theory,'' \emph{Transportation Research Record}, vol. 1883, no.~1, pp. 3--13, 2004.

\bibitem{giorgi2002traffic}
F.~Giorgi, L.~Leclercq, and J.-B. Lesort, ``A traffic flow model for urban traffic analysis: extensions of the lwr model for urban and environmental applications,'' in \emph{Transportation and Traffic Theory in the 21st Century}.\hskip 1em plus 0.5em minus 0.4em\relax Emerald Group Publishing Limited, 2002, pp. 393--415.

\bibitem{lattanzio2011moving}
C.~Lattanzio, A.~Maurizi, and B.~Piccoli, ``Moving bottlenecks in car traffic flow: a pde-ode coupled model,'' \emph{SIAM Journal on Mathematical Analysis}, vol.~43, no.~1, pp. 50--67, 2011.

\bibitem{delle2014scalar}
M.~L. Delle~Monache and P.~Goatin, ``Scalar conservation laws with moving constraints arising in traffic flow modeling: an existence result,'' \emph{Journal of Differential equations}, vol. 257, no.~11, pp. 4015--4029, 2014.

\bibitem{delle2017stability}
{Delle Monache, Maria Laura and Goatin, Paola}, ``Stability estimates for scalar conservation laws with moving flux constraints,'' \emph{Networks and Heterogeneous Media}, vol.~12, no.~2, pp. 245--258, 2017.

\bibitem{liard2019well}
T.~Liard and B.~Piccoli, ``Well-posedness for scalar conservation laws with moving flux constraints,'' \emph{SIAM Journal on Applied Mathematics}, vol.~79, no.~2, pp. 641--667, 2019.

\bibitem{liard2021entropic}
{Liard, Thibault and Piccoli, Benedetto}, ``On entropic solutions to conservation laws coupled with moving bottlenecks,'' \emph{Communications in mathematical sciences}, 2021.

\bibitem{LW55}
M.~J. Lighthill and G.~B. Whitham, ``On kinematic waves. {II}. {A} theory of traffic flow on long crowded roads,'' \emph{Proc. Roy. Soc. London Ser. A}, vol. 229, pp. 317--346, 1955.

\bibitem{R56}
P.~I. Richards, ``Shock waves on the highway,'' \emph{Operations Research}, vol.~4, pp. 42--51, 1956.

\bibitem{matin2023existence}
H.~N.~Z. Matin and M.~L.~D. Monache, ``On the existence of solution of conservation law with moving bottleneck and discontinuity in flux,'' \emph{arXiv preprint arXiv:2310.00537}, 2023.

\bibitem{temple1982global}
B.~Temple, ``Global solution of the cauchy problem for a class of 2$\times$ 2 nonstrictly hyperbolic conservation laws,'' \emph{Advances in Applied Mathematics}, vol.~3, no.~3, pp. 335--375, 1982.

\bibitem{JangReinforcement}
K.~Jang, N.~Lichtl\'e, E.~Vinitsky, A.~Shay, M.~Bunting, M.~Nice, B.~Piccoli, B.~Seibold, D.~B. Work, M.~L. Delle~Monache, J.~Sprinkle, J.~W. Lee, and A.~M. Bayen, ``Large-scale wave dampening via single-agent reinforcement learning algorithms: Controlling and deploying 100 vehicles on a highway with learning-based algorithms,'' \emph{IEEE Control Systems Magazine}, 2024.

\bibitem{lichtle2022deploying}
\BIBentryALTinterwordspacing
N.~Lichtl{\'e}, E.~Vinitsky, M.~Nice, B.~Seibold, D.~Work, and A.~M. Bayen, ``Deploying traffic smoothing cruise controllers learned from trajectory data,'' in \emph{2022 International Conference on Robotics and Automation (ICRA)}.\hskip 1em plus 0.5em minus 0.4em\relax IEEE, 2022, pp. 2884--2890. [Online]. Available: \url{https://ieeexplore.ieee.org/abstract/document/9811912}
\BIBentrySTDinterwordspacing

\bibitem{treiber2000congested}
M.~Treiber, A.~Hennecke, and D.~Helbing, ``Congested traffic states in empirical observations and microscopic simulations,'' \emph{Physical review E}, vol.~62, no.~2, p. 1805, 2000.

\bibitem{LeeMegacontrollerCSM}
J.~W. Lee, H.~Wang, K.~Jang, A.~Hayat, M.~Bunting, A.~Alanqary, W.~Barbour, Z.~Fu, X.~Gong, G.~Gunter, S.~Hornstein, A.~R. Kreidieh, N.~Lictl\'e, M.~Nice, W.~A. Richardson, A.~Shah, E.~Vinitsky, F.~Wu, X.~Shengquan, S.~Almatrudi, F.~Althukair, R.~Bhadani, J.~Carpio, R.~Chekroun, E.~Cheng, M.~T. Chiri, F.-C. Chou, R.~DeLorenzo, M.~Gibson, D.~Gloudemans, A.~Gollakota, J.~Ji, A.~Keimer, N.~Khoudari, M.~Mahmood, M.~Mahmood, H.~N.~Z. Matin, S.~McQuade, R.~Ramadan, D.~Urieli, Y.~Wang, R.~Xu, M.~Yao, Y.~You, G.~Zachar, Y.~Zhao, M.~N. Baig, S.~Bhaskaran, K.~Butts, M.~Gowda, J.~Lee, L.~Pedersen, Z.~Zhang, C.~Zhou, D.~B. Work, B.~Seibold, J.~Sprinkle, B.~Piccoli, M.~L. Delle~Monache, and A.~M. Bayen, ``Traffic smoothing via connected \& automated vehicles: A modular, hierarchical control design deployed in a 100-cav flow smoothing experiment,'' \emph{IEEE Control Systems Magazine}, 2024.

\bibitem{gloudemans202324}
D.~Gloudemans, Y.~Wang, J.~Ji, G.~Zachar, W.~Barbour, E.~Hall, M.~Cebelak, L.~Smith, and D.~B. Work, ``I-24 motion: An instrument for freeway traffic science,'' \emph{Transportation Research Part C: Emerging Technologies}, vol. 155, p. 104311, 2023.

\bibitem{CIRCLES}
``Circles,'' \url{https://circles-consortium.github.io/}, 2023.

\bibitem{gloudemans2020interstate}
D.~Gloudemans, W.~Barbour, N.~Gloudemans, M.~Neuendorf, B.~Freeze, S.~ElSaid, and D.~B. Work, ``Interstate-24 motion: Closing the loop on smart mobility,'' in \emph{2020 IEEE Workshop on Design Automation for CPS and IoT (DESTION)}.\hskip 1em plus 0.5em minus 0.4em\relax IEEE, 2020, pp. 49--55.

\bibitem{9922471}
Y.~Zhang, M.~Quinones-Grueiro, W.~Barbour, C.~Weston, G.~Biswas, and D.~Work, ``Quantifying the impact of driver compliance on the effectiveness of variable speed limits and lane control systems,'' in \emph{2022 IEEE 25th International Conference on Intelligent Transportation Systems (ITSC)}, 2022, pp. 3638--3644.

\bibitem{10207650}
Y.~Zhang, M.~Quinones-Grueiro, W.~Barbour, Z.~Zhang, J.~Scherer, G.~Biswas, and D.~Work, ``Cooperative multi-agent reinforcement learning for large scale variable speed limit control,'' in \emph{2023 IEEE International Conference on Smart Computing (SMARTCOMP)}, 2023, pp. 149--156.

\bibitem{kardous2022rigorous}
N.~Kardous, A.~Hayat, S.~T. McQuade, X.~Gong, S.~Truong, T.~Mezair, P.~Arnold, R.~Delorenzo, A.~Bayen, and B.~Piccoli, ``A rigorous multi-population multi-lane hybrid traffic model and its mean-field limit for dissipation of waves via autonomous vehicles,'' \emph{arXiv preprint arXiv:2205.06913}, 2022.

\bibitem{hayat2022holistic}
A.~Hayat, X.~Gong, J.~Lee, S.~Truong, S.~McQuade, N.~Kardous, A.~Keimer, Y.~You, S.~Albeaik, E.~Vinistky, P.~Arnold, M.~L. Delle~Monache, A.~Bayen, B.~Seibold, J.~Sprinkle, D.~Work, and B.~Piccoli, \emph{A Holistic Approach to the Energy-Efficient Smoothing of Traffic via Autonomous Vehicles}.\hskip 1em plus 0.5em minus 0.4em\relax Cham: Springer International Publishing, 2022, pp. 285--316.

\bibitem{lee2021integrated}
\BIBentryALTinterwordspacing
J.~W. Lee, G.~Gunter, R.~Ramadan, S.~Almatrudi, P.~Arnold, J.~Aquino, W.~Barbour, R.~Bhadani, J.~Carpio, F.-C. Chou, M.~Gibson, X.~Gong, A.~Hayat, N.~Khoudari, A.~R. Kreidieh, M.~Kumar, N.~Lichtl\'{e}, S.~McQuade, B.~Nguyen, M.~Ross, S.~Truong, E.~Vinitsky, Y.~Zhao, J.~Sprinkle, B.~Piccoli, A.~M. Bayen, D.~B. Work, and B.~Seibold, ``Integrated framework of vehicle dynamics, instabilities, energy models, and sparse flow smoothing controllers,'' in \emph{Proceedings of the Workshop on Data-Driven and Intelligent Cyber-Physical Systems}, ser. DI-CPS'21.\hskip 1em plus 0.5em minus 0.4em\relax New York, NY, USA: Association for Computing Machinery, 2021, p. 41–47. [Online]. Available: \url{https://doi.org/10.1145/3459609.3460530}
\BIBentrySTDinterwordspacing

\bibitem{LeeMegacontroller}
J.~W. Lee, H.~Wang, K.~Jang, A.~Hayat, M.~Bunting, A.~Alanqary, W.~Barbour, Z.~Fu, X.~Gong, G.~Gunter, S.~Hornstein, A.~R. Kreidieh, N.~Lictl\'e, M.~Nice, W.~A. Richardson, A.~Shah, E.~Vinitsky, F.~Wu, X.~Shengquan, S.~Almatrudi, F.~Althukair, R.~Bhadani, J.~Carpio, R.~Chekroun, E.~Cheng, M.~T. Chiri, F.-C. Chou, R.~DeLorenzo, M.~Gibson, D.~Gloudemans, A.~Gollakota, J.~Ji, A.~Keimer, N.~Khoudari, M.~Mahmood, M.~Mahmood, H.~N.~Z. Matin, S.~McQuade, R.~Ramadan, D.~Urieli, Y.~Wang, R.~Xu, M.~Yao, Y.~You, G.~Zachar, Y.~Zhao, M.~N. Baig, S.~Bhaskaran, K.~Butts, M.~Gowda, J.~Lee, L.~Pedersen, Z.~Zhang, C.~Zhou, D.~B. Work, B.~Seibold, J.~Sprinkle, B.~Piccoli, M.~L. Delle~Monache, and A.~M. Bayen, ``Traffic smoothing via connected \& automated vehicles: A modular, hierarchical control design deployed in a 100-cav flow smoothing experiment,'' \emph{IEEE Control Systems Magazine}, 2024.

\bibitem{fu2023massive}
\BIBentryALTinterwordspacing
Z.~Fu, K.~Manke, and A.~M. Bayen, ``Massive {CAV} experiment in nashville pits machine learning against traffic jams,'' apr 2023. [Online]. Available: \url{https://doi.org/10.1287%2Forms.2023.02.05}
\BIBentrySTDinterwordspacing

\bibitem{AmeliLivetraffic}
M.~Ameli, S.~McQuade, J.~W. Lee, R.~DeLorenzo, M.~L. Delle~Monache, S.~Hornstein, M.~Mahmood, M.~Mahmood, M.~Nice, M.~Bunting, D.~Timsit, R.~Wagner, R.~Xu, B.~Seibold, D.~B. Work, J.~Sprinkle, B.~Piccoli, and A.~M. Bayen, ``Designing, simulating, and performing the 100-av field test for the circles consortium: Methodology and implementation of the largest mobile traffic control experiment to date,'' \emph{IEEE Control Systems Magazine}, 2024.

\bibitem{gloudemans2023so}
D.~Gloudemans, G.~Zach{\'a}r, Y.~Wang, J.~Ji, M.~Nice, M.~Bunting, W.~Barbour, J.~Sprinkle, B.~Piccoli, M.~L. Monache, A.~Bayen, B.~Seibold, and D.~B. Work, ``So you think you can track?'' \emph{arXiv preprint arXiv:2309.07268}, 2023.

\bibitem{gloudemans2023interstate}
D.~Gloudemans, Y.~Wang, G.~Gumm, W.~Barbour, and D.~B. Work, ``The interstate-24 3d dataset: a new benchmark for 3d multi-camera vehicle tracking,'' \emph{arXiv preprint arXiv:2308.14833}, 2023.

\bibitem{gloudemans2021vehicle}
D.~Gloudemans and D.~B. Work, ``Vehicle tracking with crop-based detection,'' in \emph{2021 20th IEEE International Conference on Machine Learning and Applications (ICMLA)}.\hskip 1em plus 0.5em minus 0.4em\relax IEEE, 2021, pp. 312--319.

\bibitem{wang2023onlinemcf}
Y.~Wang, J.~Ji, W.~Barbour, and D.~B. Work, ``Online min cost circulation for multi-object-tracking on fragments,'' in \emph{2023 IEEE International Intelligent Transportation Systems Conference (ITSC)}.\hskip 1em plus 0.5em minus 0.4em\relax IEEE, 2023, to appear.

\bibitem{wang2022automatic}
Y.~Wang, D.~Gloudemans, Z.~N. Teoh, L.~Liu, G.~Zach{\'a}r, W.~Barbour, and D.~Work, ``Automatic vehicle trajectory data reconstruction at scale,'' \emph{arXiv preprint arXiv:2212.07907}, 2022.

\bibitem{10.1145/3576914.3587710}
\BIBentryALTinterwordspacing
G.~Zach\'{a}r, ``Visualization of large-scale trajectory datasets,'' ser. CPS-IoT Week '23.\hskip 1em plus 0.5em minus 0.4em\relax New York, NY, USA: Association for Computing Machinery, 2023, p. 152–157. [Online]. Available: \url{https://doi.org/10.1145/3576914.3587710}
\BIBentrySTDinterwordspacing

\bibitem{edie1963discussion}
L.~C. Edie \emph{et~al.}, \emph{Discussion of traffic stream measurements and definitions}.\hskip 1em plus 0.5em minus 0.4em\relax Port of New York Authority New York, 1963.

\end{thebibliography}

\endarticle

\end{document}